\documentclass{sig-alternate}
\usepackage{amssymb}
\usepackage{graphicx}
\usepackage{amsmath}
\usepackage{subfigure}

\begin{document}

\title{From User Comments to On-line Conversations}

%\numberofauthors{3}
\author{
\begin{tabular}{c@{\hspace*{2em}}c@{\hspace*{2em}}c}
Chunyan Wang$^\star$                & Mao Ye$^{\ddagger} $          & Bernardo A. Huberman$^\ddagger$\\
\texttt{chunyan@stanford.edu}    & \texttt{mxy177@cse.psu.edu}  & \texttt{bernardo.huberman@hp.com}\\
\end{tabular}\vspace{0.25cm}\\
{\footnotesize $^\star$Department of Applied Physics, Stanford University, California, USA}\vspace{0.1cm}\\
{\footnotesize $^\ddagger$Social Computing Group, HP Labs, California, USA}\vspace{0.1cm}\\
} 
%\date{}
\maketitle

\begin{abstract}
We present an analysis of user conversations in on-line social media and their evolution over time. We propose a dynamic model that accurately predicts the growth dynamics and structural properties of conversation threads. The model successfully reconciles the differing observations that have been reported in existing studies. By separating artificial factors from user behaviors, we show that there are actually underlying rules in common for on-line conversations in different social media websites. Results of our model are supported by empirical measurements throughout a number of different social media websites.
\end{abstract}

% A category with the (minimum) three required fields % to be revised
%\category{J.4}{Computer Applications}{Social and Behavior Sciences}
%\category{G.3}{Mathematics of Computing}{Probability and Statistics}
%\terms{Human Factors,Theory, Measurement}

\section{Introduction}
The rapid development of social media websites has dramatically changed the way that people communicate with each other.  A particularly interesting phenomenon is the prominent role of users as a leading information source within these websites. For example, various on-line media and review sites provide commenting facilities for users to exchange opinions and express sentiments about news, stories and products. These user-generated comments link together and form a conversation thread, which is essentially a distinctive kind of information network that has a life span significantly shorter than other information networks such as forums and other on-line communities. As pointed out in~\cite{K10}, despite the significant research on the dynamics of networks of linked information, networks like conversation threads have not received enough attention so far. In fact, the dynamics of conversations plays a fundamental role in opinion spread and formation~\cite{G05,W07}, word-of-mouth effects~\cite{B09} and collective problem solving~\cite{Gr05,K06}. Existing empirical studies on on-line conversations seem to yield conflicting results about the basic statistical properties. Some of existing studies demonstrate that the size distribution of posts and reviews follow a heavy-tailed distribution such as Zipf's law~\cite{M06, K10} or log-normal distribution~\cite{K11,G08}, another portion of the literature suggest a light-tailed one, such as negative binomial distribution~\cite{O08,T09}. A fundamental question is how can two apparently different categories of distributions describe the same type of information network? And what are the dominating factors that are responsible for the observed differences? In this paper, we focus on addressing these problems by proposing a dynamic model for on-line conversations. Some of our key findings are summarized below.

\begin{itemize}
\item{{\bf User Attention on New Items.}  We examine the dynamics of user attention on new items since their creation. We analyze the duration of new topics displayed to users. We also investigate the non-Poisson nature of user commenting behavior.} 

\item{{\bf Model of On-line Conversations.}  We propose a dynamic model for conversation growth based on a number of factors, including the exposure duration of topics on the website, the patterns of user commenting behavior, the interestingness of topics and the impacts of social propagation and resonance. The model successfully reconciles existing discrepancies in reported studies. We also extend the model to explain the structural properties within a conversation thread.}

\item{{\bf Size and Structure of Conversations.} We compare results from our model with empirical measurements using datasets from Digg\footnote{http://digg.com}, Reddit\footnote{http://reddit.com} and Epinions\footnote{http://epinions.com}.}
\end{itemize}

The rest of this paper is organized as follows. Section 2 reviews the related work. Section 3 introduces the datasets used in our empirical studies. Section 4 introduces key observations on user behavior and the impact of featuring mechanisms on user attention. Based on these observations, Section 5 introduces the dynamic model for user conversation. Section 6 compares predictions of the model with empirical measurements. Section 7 concludes the paper with a discussion.

\section{Related Work}

\subsection{Information Spread and Conversations}
There has been significant research on information dissemination. The pioneering work of Liben-Nowell and Kleinberg~\cite{L08} modeled information spread as a propagation of chain letter. Golub and Jackson~\cite{G10} extend this work with a branching process model combined with the selection bias of observation. In social media, Leskovec et al.~\cite{J09} investigated the propagation of memes across the Web. The main concern of these studies is understanding the mechanism of information spread in the context of social network. Others focus on properties of information networks, such as on-line conversations, that are formed in the process of information spread. Mishne et al.~\cite{M06} looked at web-log comments for identifying blog post controversy, Duarte et al.~\cite{D07} engaged in describing blogsphere access patterns from the blog server point, Kaltenbrunner et al.~\cite{K11} measured community response time in terms of comment activity on Slashdot stories, Choudhury et al.~\cite{Ch09} characterize conversations through their interestingness, and finally, Kumar et al. modeled the dynamics of conversations with a branching process incorporating recency. Despite of the increasing interests in on-line conversations, one problem not addressed so far is the seemingly conflicting observations about on-line conversation's basic statistical properties. As mentioned earlier, some studies suggest that the size follows a heavy-tailed distribution\footnote{In this paper, we use the term heavy-tailed to denote the probability distributions whose tails are not exponentially bounded, i.e. $\mathop {\lim }\limits_{x \to \infty } {e^{\lambda x}}P(X > x) = \infty ,\lambda  > 0$.}  such as Zipf's law~\cite{M06, K10} or log-normal distribution~\cite{K11,G08}, other measurements point to a Poisson family~\cite{O08,T09}. One main focus of this paper is to provide an explanation for these differing observations.

\subsection{Dynamics of User Attention}
Another set of studies related to our work is the dynamics of user attention. Along with the outbreak of information, topics on websites compete with each other for the scarce attention of users~\cite{B01,F08}. To help users find high quality content, social media websites usually place information in a ``featured column'' for popular items, such as the ``popular'' page on Youtube, the ``trending'' column on Twitter, the ``what's hot'' column on Reddit. Existing studies show that this featuring mechanism has significant impact on user attention~\cite{E08}. To enter the featured column, topics need to reach a threshold of critical mass. For instance, studies on Digg~\cite{F07,L10}, Youtube~\cite{C08}, Wikipedia~\cite{R10} and Twitter~\cite{B09,A11} successfully explain the attention dynamics of topics after the critical mass threshold. However, it is still unclear about the attention dynamics of the vast majority of topics and stories that never reach the critical mass. As such, it has remained an open question about the attention dynamics and the initial growth of these items. We attempt to propose dynamics of the user attention, measured in the number of user comments, for these general items on social media websites.

\subsection{Bursty Nature of Waiting Times}
Recent advances in technology have made possible the study of human dynamics, one of which subject aims to address the timing of many human activities. In contrast with the traditional framework, which describes waiting times under the context of Internet as a series of Poisson processes~\cite{S03}, recent observations from data on email exchanges ~\cite{B05,E04,R09} and web browsing~\cite{H98,C08,E09,A10} suggest that the waiting times for human activities follows power law scaling. Various models have been proposed to interpret the observed bursts of waiting times~\cite{B05, A10, J10}. Most existing studies emphasize on explaining the nature and origin of the bursts, rather than explore the implications of this observation on information spread and attention dynamics. In this paper, we examine the waiting times of human comments and more importantly, we extend the study by using this observation of human behavior to explain the dynamics of user attention and the growth of on-line conversations.

\section{Data}
Three datasets from Digg, Reddit and Epinions are used in our empirical measurements. To collect these datasets, we monitored the website for newly created items or topics. We kept track of these topics' user comments for a time span of at least three months since the topics' creation, to make sure that the growth saturates. We also recorded related information such as the time stamp when the topic is removed from the column for displaying new items. In our empirical studies, we perform the same treatments on these datasets whenever possible.

Digg is an interactive social media website, which allows its users to share and comment on news and stories. Users of the website select and direct
attention to a few items from a very large pool of submissions. They can read, Digg, Bury, and leave comments on the topic or other users' comments. In our study, we monitored the website for a total number of $17,322$ topics containing $158,782$ comments. Each comment was labeled by its posting time. To obtain information about individual user's commenting behavior, we also monitored a number of $8,616$ users on Digg and collected all these users' comments. 

Another dataset used in our study was from the social news website Reddit. Users on Reddit submit content in the form of either a link or a text post. Other users reading the post can express their opinions by commenting on the original post. Similar to Digg, comments on Reddit can also be directed to existing comments. In our study, we collected over $78,312$ comments from $8,428$ conversation threads. For each comment, we recorded the user-id and timestamps of the comments. We also recorded to which each comment is referring to. User information 

To ensure that our observations are not limited to news media sites, we included a dataset of consumer review from Epinions. Epinions is a who-trust-whom consumer review site, and users write their personal reviews on a wide variety of products, ranging from automobiles to media (music, books, movies and etc.). Members of the site can decide whether to trust other members based on their reviews. Again, every user on the website can comment on the reviews or on the existing comments. We collected $88,859$ unique users' comments from the website. And also $286,317$ topics containing $722,475$ user comments.

%Table 1 gives a summarization of the datasets.
%
%\begin{table}[htl]
%\centering
%\begin{tabular}{|l|c|c|c|}
%\hline & Comments & Threads\\
%\hline Digg & 158,782 & 17,322 \\
%\hline Reddit & 78,312 & 8,428 \\
%\hline Epinions  & 810,527 & 286,317 \\
%\hline
%\end{tabular}
%\caption{Synopsis of the datasets.}\label{tbl:datasets}
%\end{table}

\section{USER ATTENTION TO NEW ITEMS}

\begin{figure}[htl]
 \centering
\hspace{-10pt}
  \subfigure[Cumulative Size]{\includegraphics[width=4.5cm,height=3.4cm]{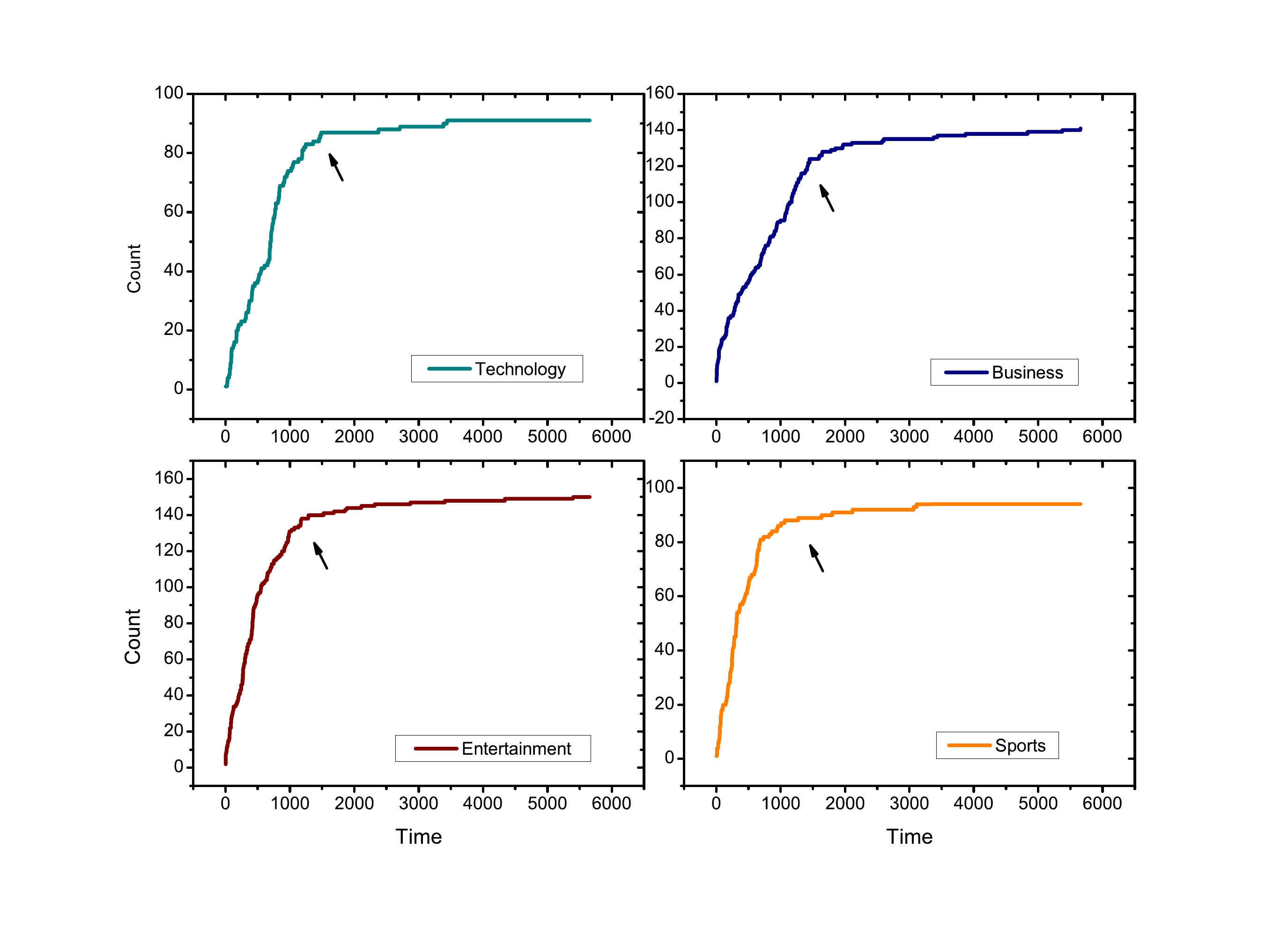}}
\hspace{-15pt}
\subfigure[Ratio]{\includegraphics[width=4.5cm,height=3.50cm]{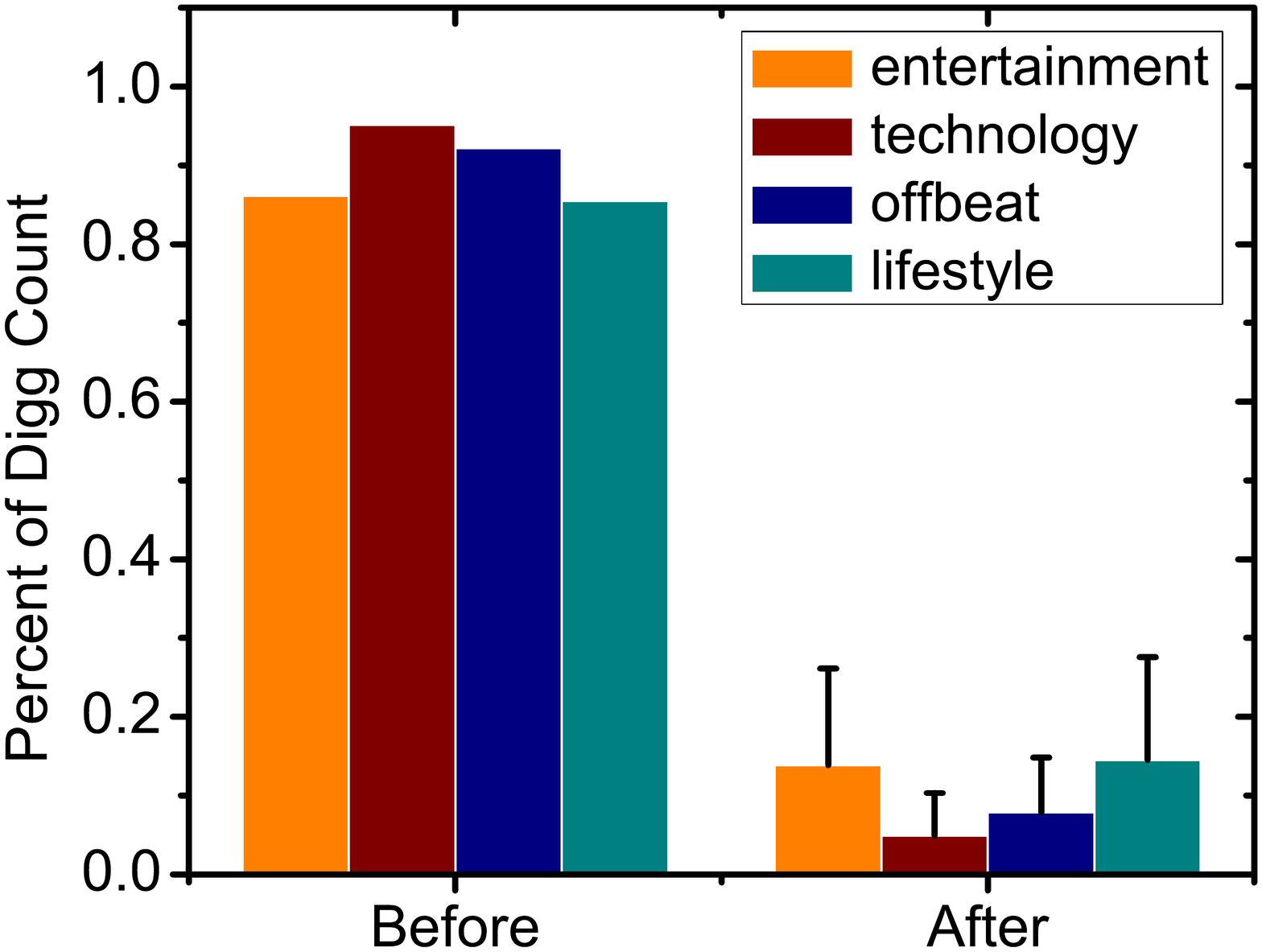}}
\hspace{-10pt}
 \caption{(a) User attention as a function of time (minutes) for four typical topics on Digg. The black arrow in each plot shows the inflection point where the topic is removed from the``upcoming'' column. (b) Percentage of user comments happened before and after the inflection point. }
 \label{fig:Digg_time}
\end{figure}

To understand the underlying mechanisms governing user attention and on-line conversations, we first look at the growth of attention on newly generated items in social media. Figure~\ref{fig:Digg_time} (a) shows the growth of cumulative user attention measured in Digg count of four typical topics from Digg. Results from Reddit and Epinions are similar to the one in Digg. One general observation for topics from different categories and different websites is that the cumulative count saturates to a point where a sharp drop of the growth rate is apparent. We explain this observation with the following reasoning. To help users explore new topics, typical social media websites place newly generated topics in the ``upcoming'' and ``new'' columns since their creation time. Users visit the website regularly and discover these newly generated stories. After a period of time, these old topics are replaced with newly generated contents. While the replaced item can still be accessed through search queries, it has significantly less chance to be exposed to general users. So this explains why the growth of attention eventually saturates. To confirm this explanation, we kept track of the time when the topics are removed from the front page of ``upcoming column'' on Digg. We find that the saturation point has a high correlation with the time point when the topic is removed. The black arrow in Figure~\ref{fig:Digg_time} (a) identifies the time point of removal, which is very close to the saturation point. Figure~\ref{fig:Digg_time} (b) compares the number of user comments happened before and after the inflection point. The averaged percentages of comments happened before the inflection points are $0.8616$, $0.9509$, $0.9215$ and $0.8548$ respectively for categories of entertainment, technology, offbeat and lifestyle on Digg. Different colors in the plot represent different sub-categories. Error bars in the plot indicate one standard deviation of the data in the sub-category. As expected, most of the comments are generated before the inflection point. 

In the rest of this paper, we name the time point, when a topic is replaced from ``new'' column, as the ``inflection point''. And we denote the duration that a topic stays in the column as ``exposure duration''. The exposure duration varies from topic to topic, which is largely determined by the speed of creating new items and the hidden algorithms used by the website to remove old topics. In the following of this paper, we focus on the growth dynamics before the inflection point. There are two important factors that are dominating this initial growth: (i) the length of exposure duration and (ii) the patterns of user commenting behavior.  Now we focus on studying these two factors.

\subsection{Distribution of Exposure Durations}
The duration of items placed in the ``new'' column since creation plays a fundamental role in the initial growth of attention dynamics and comment counts. Here, we empirically measure the distribution of this exposure duration from three mainstream social media websites Digg, Reddit and Epinions.
\begin{figure}[htl]
 \centering
\hspace{-10pt}
  \subfigure[Digg \& Reddit] {\includegraphics[width=4.5cm]{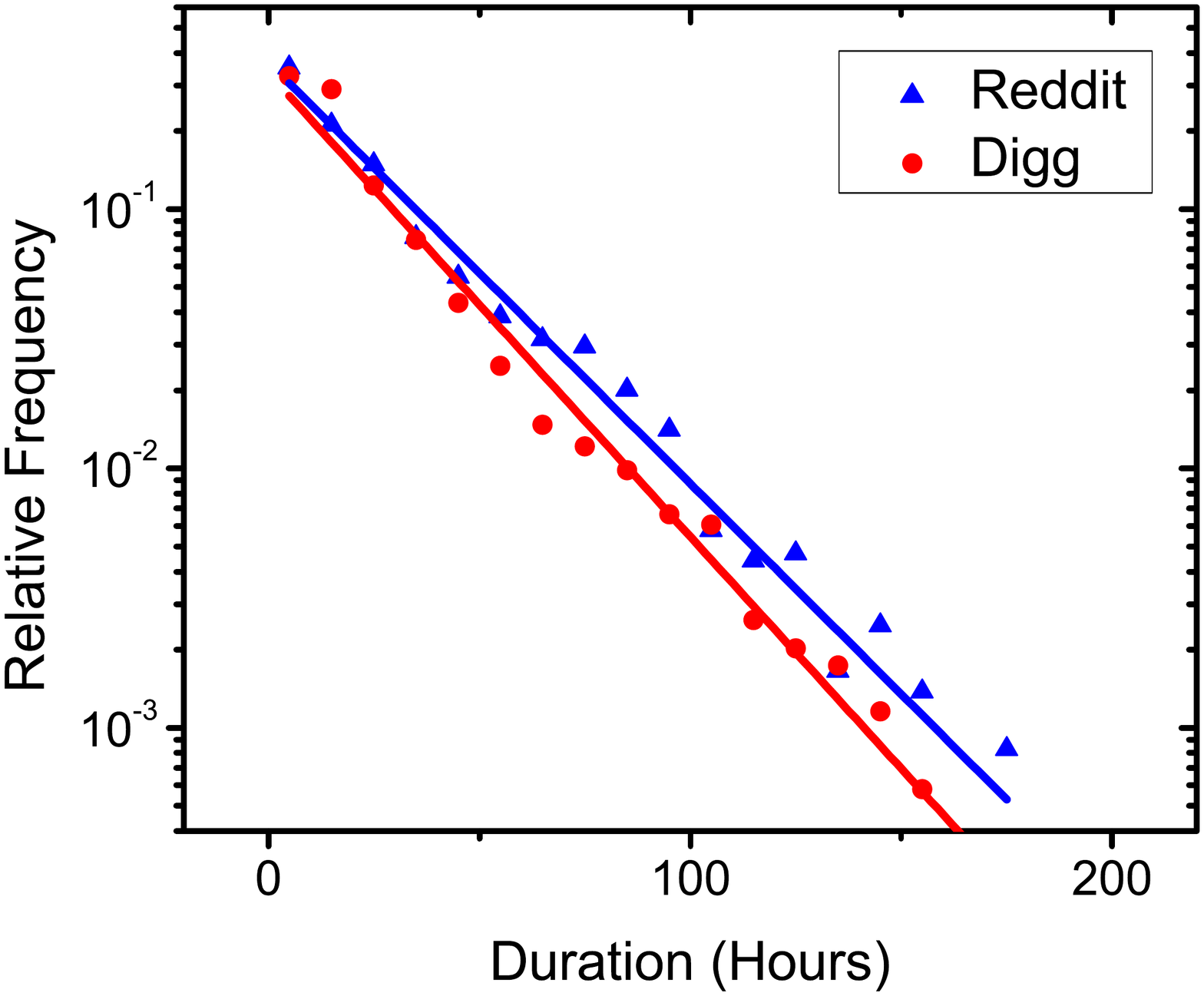}}
\hspace{-15pt}
  \subfigure[Epinions]{\includegraphics[width=4.5cm]{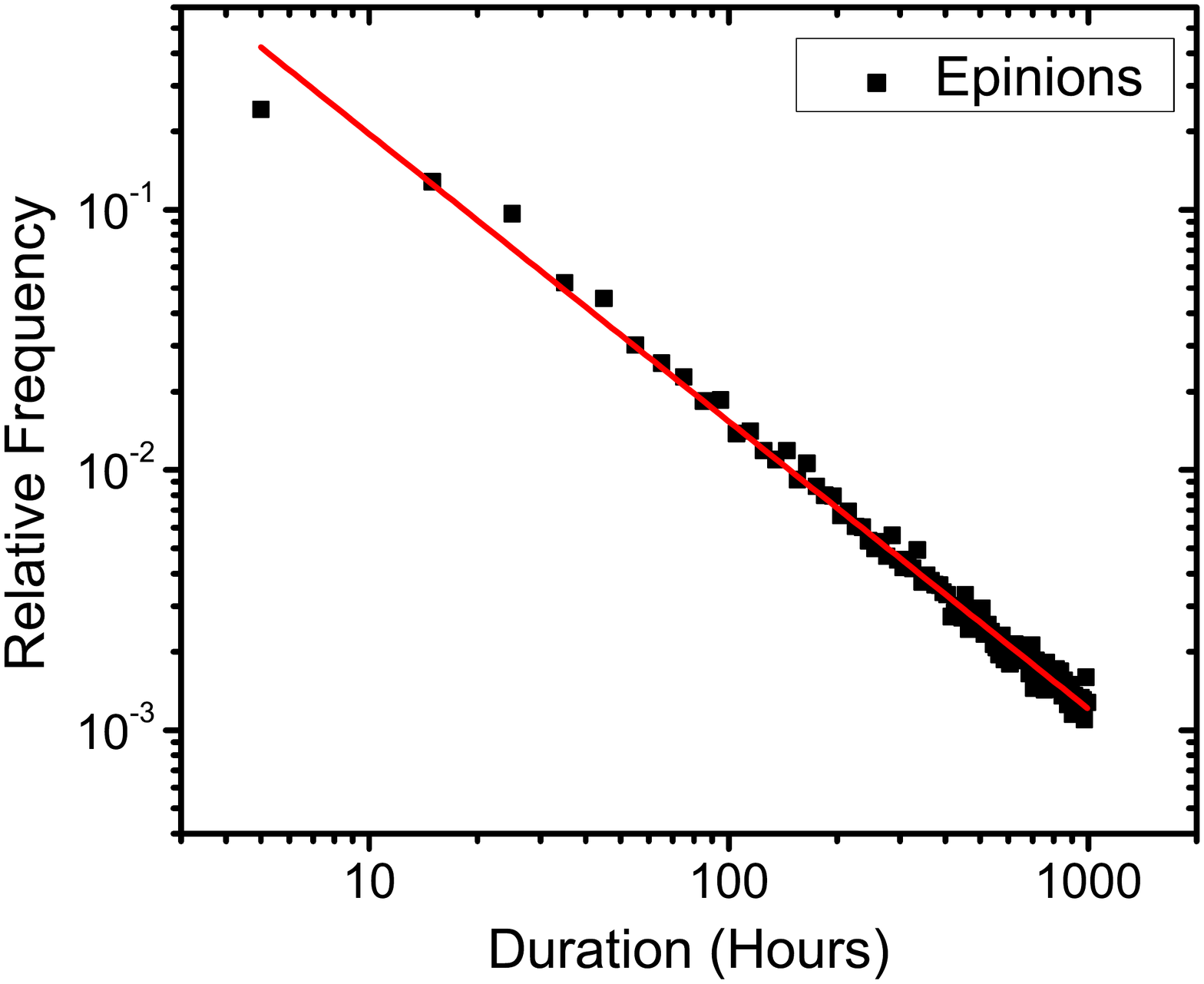}}
\hspace{-10pt}
 \caption{Density plot of exposure duration for new topics. (a) Digg and Reddit, (b) Epinions. }
 \label{fig:Topic_duration}
\end{figure}

On Digg, there is a specific column named ``upcoming news'' for newly generated items. Topics in this column are sorted by creation time, with the newest item ranking on the top of the page. When new topic comes, all of the existing items move downwards on the web page. In doing so, topics of the website would fade away from users' attention gradually. Here we measure the duration that the item maintains on the first top $50$ items in the ``upcoming news' column. Similar results are observed when we change this threshold limit. On Reddit and Epinions, we use the same gathering methodologies and treatments. In Figure~\ref{fig:Topic_duration} (a), an exponential distribution can be observed from the semi-log plot for both of Digg and Reddit. And for Epinions in Figure~\ref{fig:Topic_duration} (b), a Pareto distribution for the exposure duration is observed from the straight line in log-log plot. Since the exposure duration is determined by various specific factors such as the speed of item creation and the underlying algorithms used, it is normal to observe different distribution of exposure durations. The duration is expected to be exponentially distributed, if items in the column are removed with a fixed probability in each time step~\cite{A11}. Various optimizing strategies can result in a power law distribution or a log-normal distribution of exposure durations~\cite{He09}. For this reason, one could not presume the distribution of exposure duration without knowledge about the hidden algorithms or empirical measurements. The impact of differences in exposure duration is later discussed in the model section.

\begin{figure*}[htl]
 \centering
  \subfigure[Individuals]{\includegraphics[width=6.0cm]{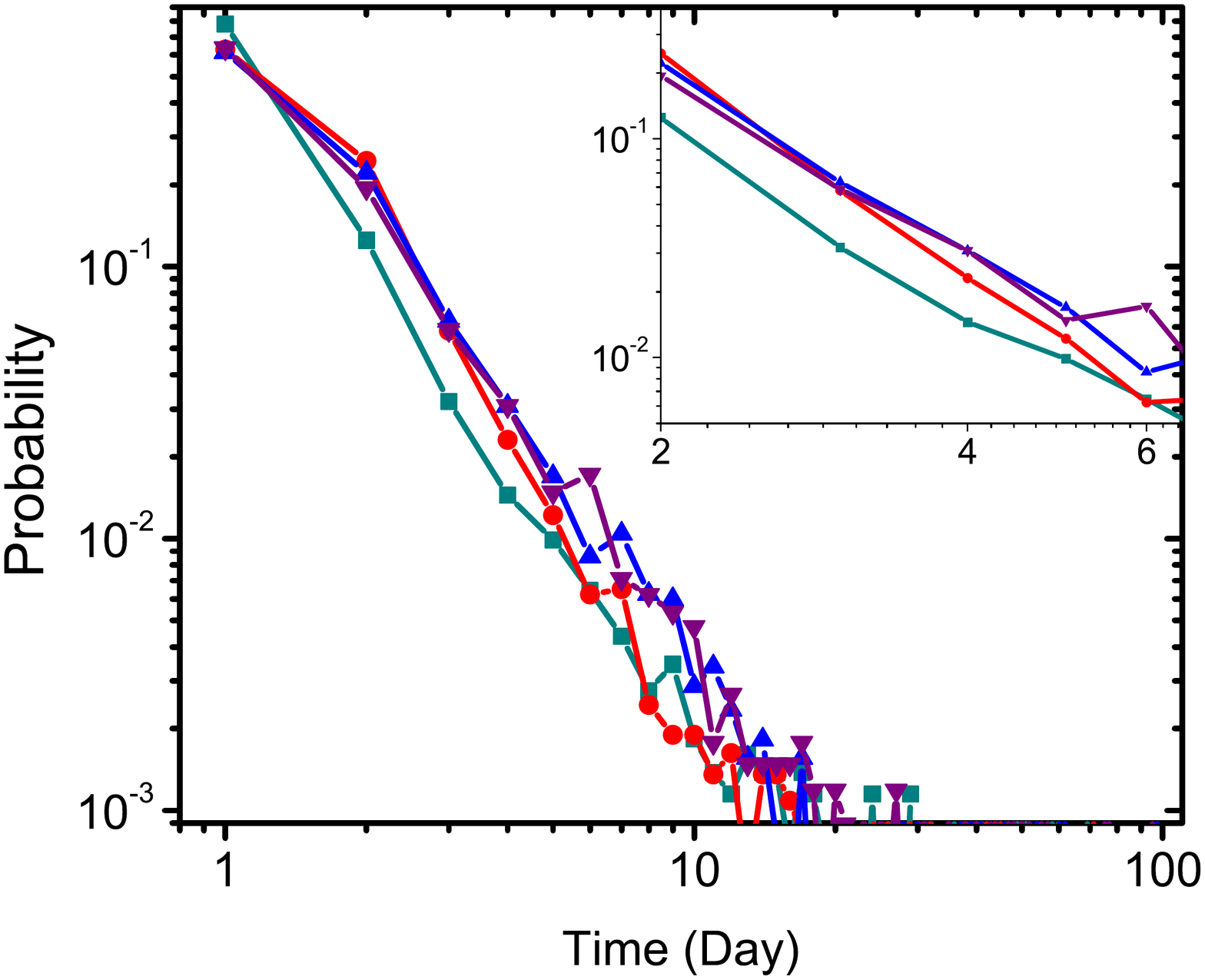}}
\hspace{-10pt}
  \subfigure[Epinions] {\includegraphics[width=6.0cm]{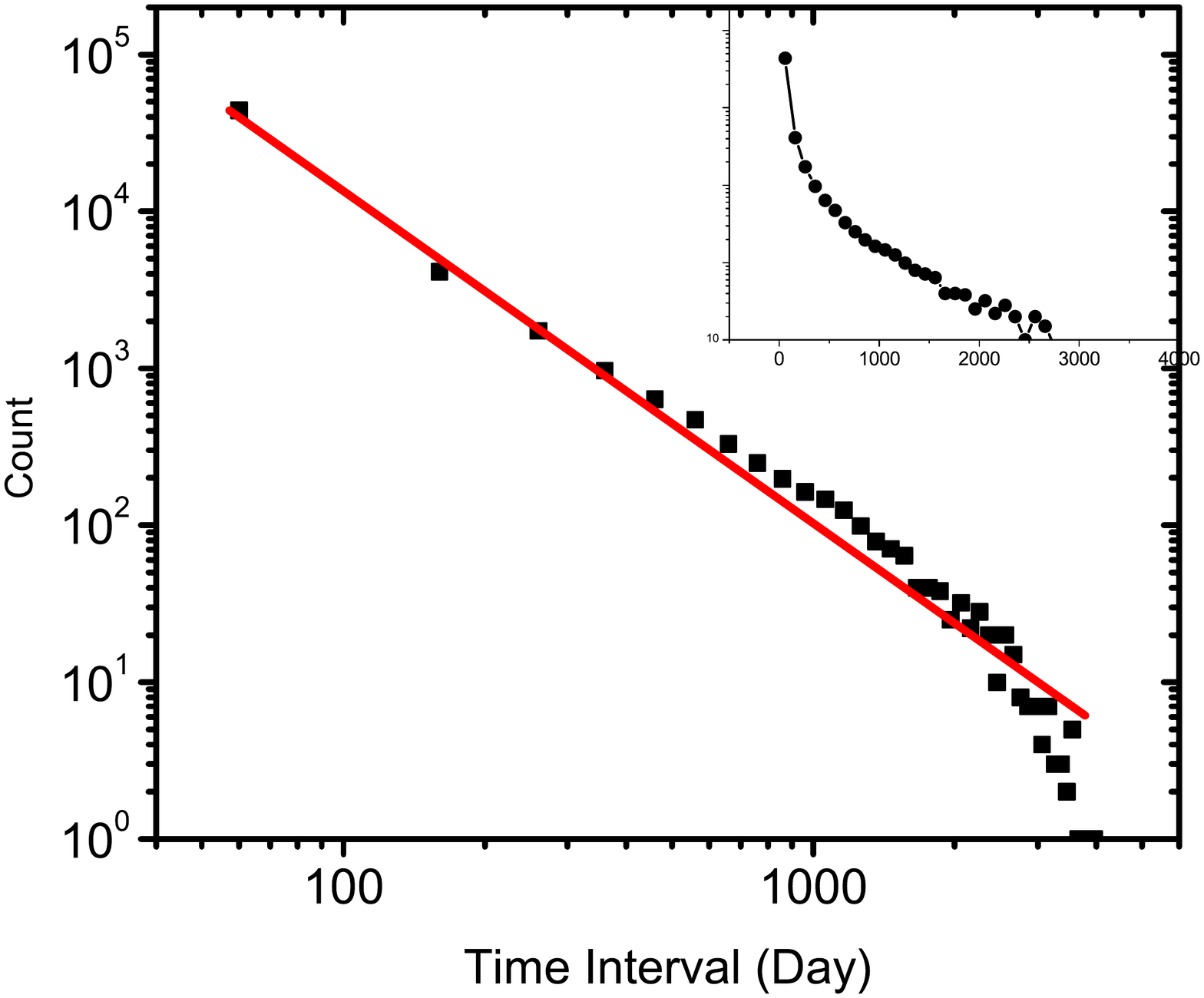}}
\hspace{-10pt}
  \subfigure[Digg]{\includegraphics[width=6.0cm]{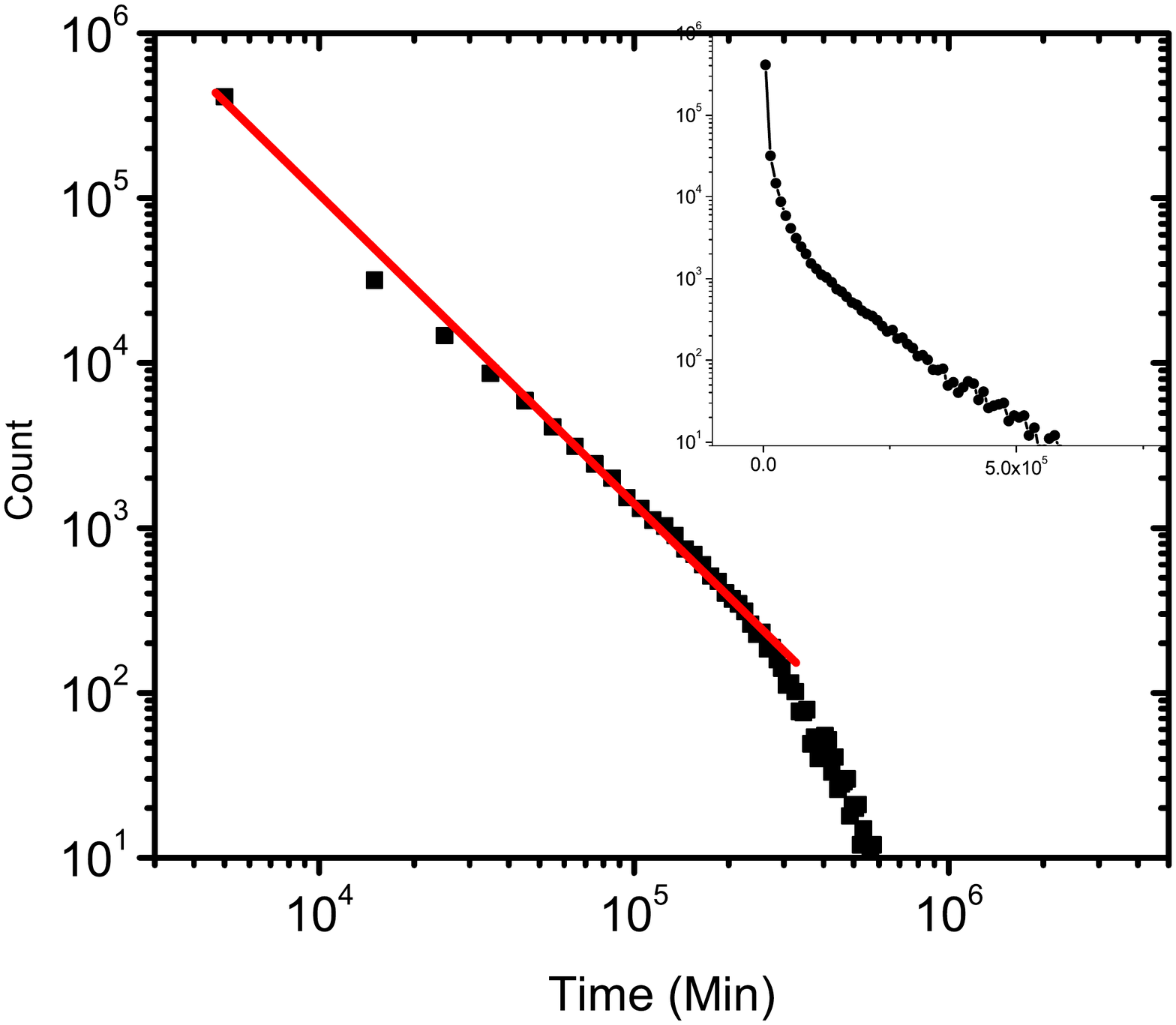}}
 \caption{Density plot of waiting times between two consecutive comments from a user, (a) time intervals of four typical users on Digg, (b) intervals from all users in Epinions dataset and (c) intervals from all users in Digg dataset. The upper right plot in (a) shows a zoomed-in view of density plot. The red straight line in (b) and (c) suggests a power-law family of distribution. The plot on the upper right corner in (b) and (c) demonstrates data in a semi-log scale.}
 \label{fig:user_comment_interval}
\end{figure*}

\subsection{Patterns of User Commenting Behavior}
In last sub-section, we focus on the side of websites, looking at the distribution of for how long they will choose to display a new item to its users. Now, we turn our attention to users' commenting behavior, i.e. the distribution of waiting times between two comments from the same user.

First, we look at the distribution of two consecutive comments from single users. Figure~\ref{fig:user_comment_interval} (a) demonstrates the distribution of waiting times for four typical users on Digg in a log-log scale. The upper right plot in the figure shows the scaling region ranging from $2$ to $6$ days. One interesting observation from the plot is that the four colored lines, despite coming from different users, show similar scaling relationship. And the slope of the line also varies little in the four samples. This is suggesting that different users share similar patterns of commenting behaviors. So we turn our attention to study the behavior of aggregated users on a whole, by treating users as identical. We empirically measure the distribution of waiting times by collecting the time series data of all comments from users. The density plot of waiting times between two consecutive comments in a log-log scale is shown in  Figure~\ref{fig:user_comment_interval} (b) and (c). The red straight line in the plot is not an actual fit of data but a guidance of eye, which suggests a power-law scaling of waiting times distribution. The plots on the upper right corner demonstrate the exact same data but in a semi-log scale. From these two plots, the distribution clearly deviates from an exponential distribution. The cutoff at around $1000$ days for (b) and $10^5$ minutes for (c) can be explained by the finite-size effect, which may stem from the limited life span of the websites. The above observations suggest that the commenting behavior of human can not be described by a Poisson process as assumed in prior studies~\cite{S03}. We find that the density plot is best fitted with a upper-truncated Pareto distribution. Based on the maximum-likelihood estimation (MLE) approach~\cite{C09} for upper-truncated Pareto distribution, the exponent for Epinions is estimated to be $-1.5670$, when the lower bound is set to equal one unit and the upper bound is set to be equal to the largest observation in our records. Similarly for the MLE of Digg dataset, the exponent is estimated to be $-1.1262$. This result implies that, for each user, frequent comments may follow by a significantly long period of inactivity. In the following, we explore the implications of this non-Poisson nature of human behavior.

\section{MODEL OF ON-LINE CONVERSATIONS}
We introduced basic properties about the duration of new topics getting displayed to users and patterns of user commenting behavior. In this section, based on these properties, we propose a model for the growth dynamics of on-line conversations. The model explains differing observations in conversation size distribution that have been reported. We also extend the model by the Yule process to explain the in-degree distribution of each comment. 

\begin{figure}[htl]
 \centering
 \includegraphics[width=6.0cm]{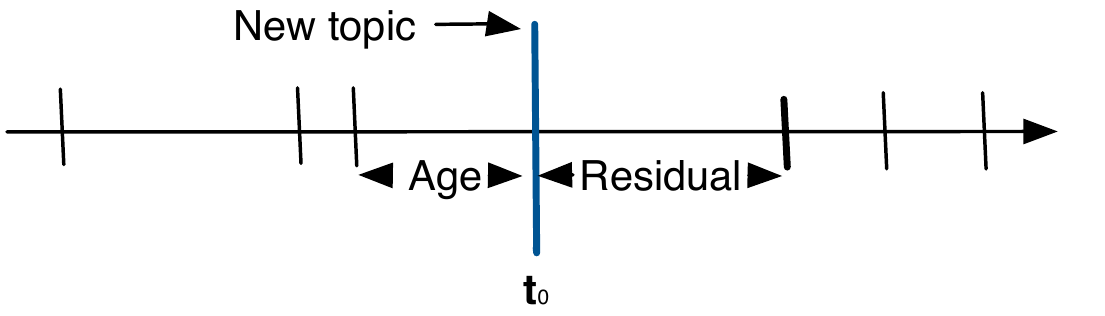}
 \caption{The arrival pattern of comments from one user. Every short vertical line in the figure represents the time of a comment from the user. The blue thick vertical bar represents the time point when a new topic is released.}
 \label{fig:renewal_process_demo}
\end{figure}

\begin{figure}[htl]
 \centering
\subfigure[Digg dN-dT] {\includegraphics[width=4.0cm]{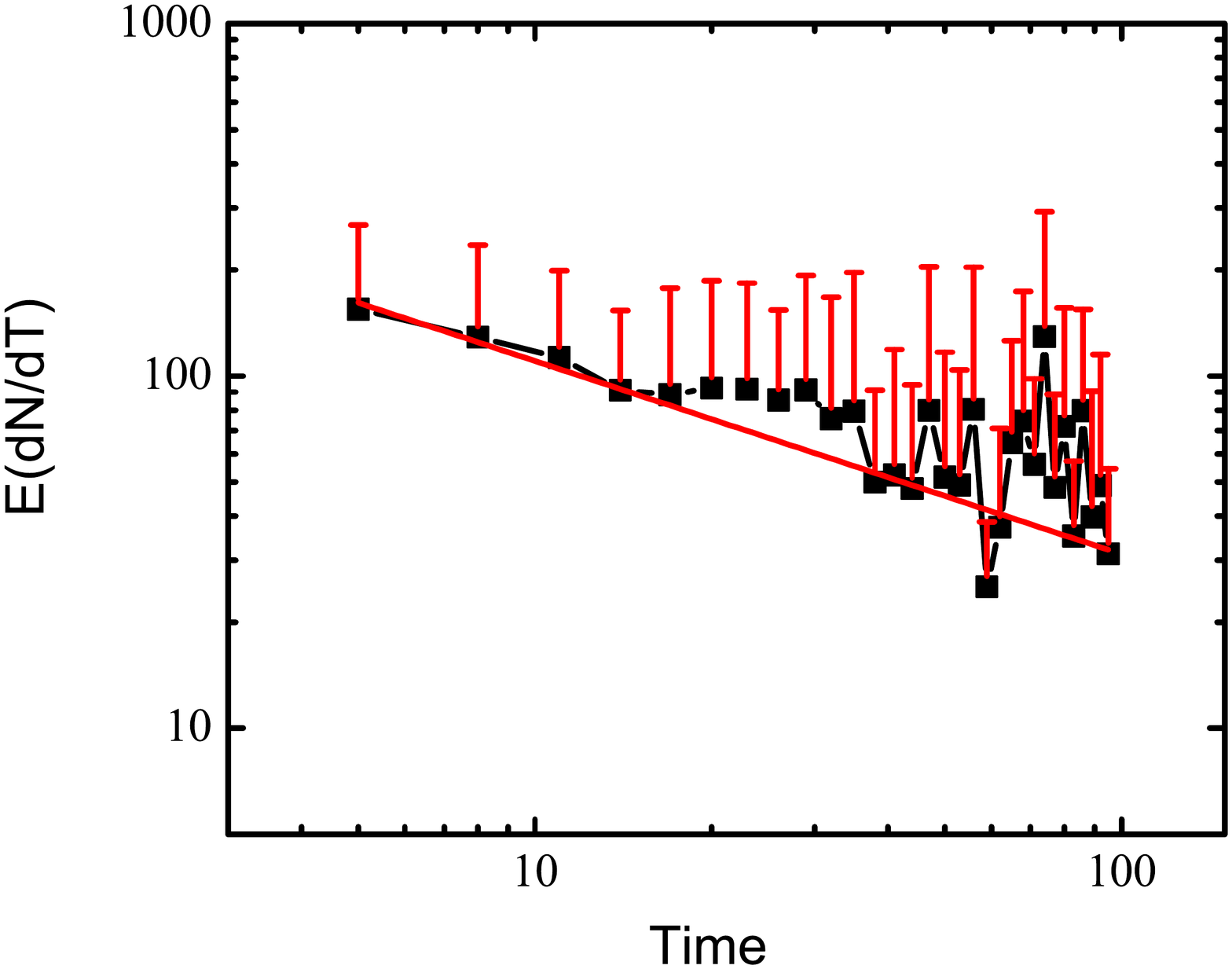}}
\subfigure[Digg N-dT]{\includegraphics[width=4.0cm]{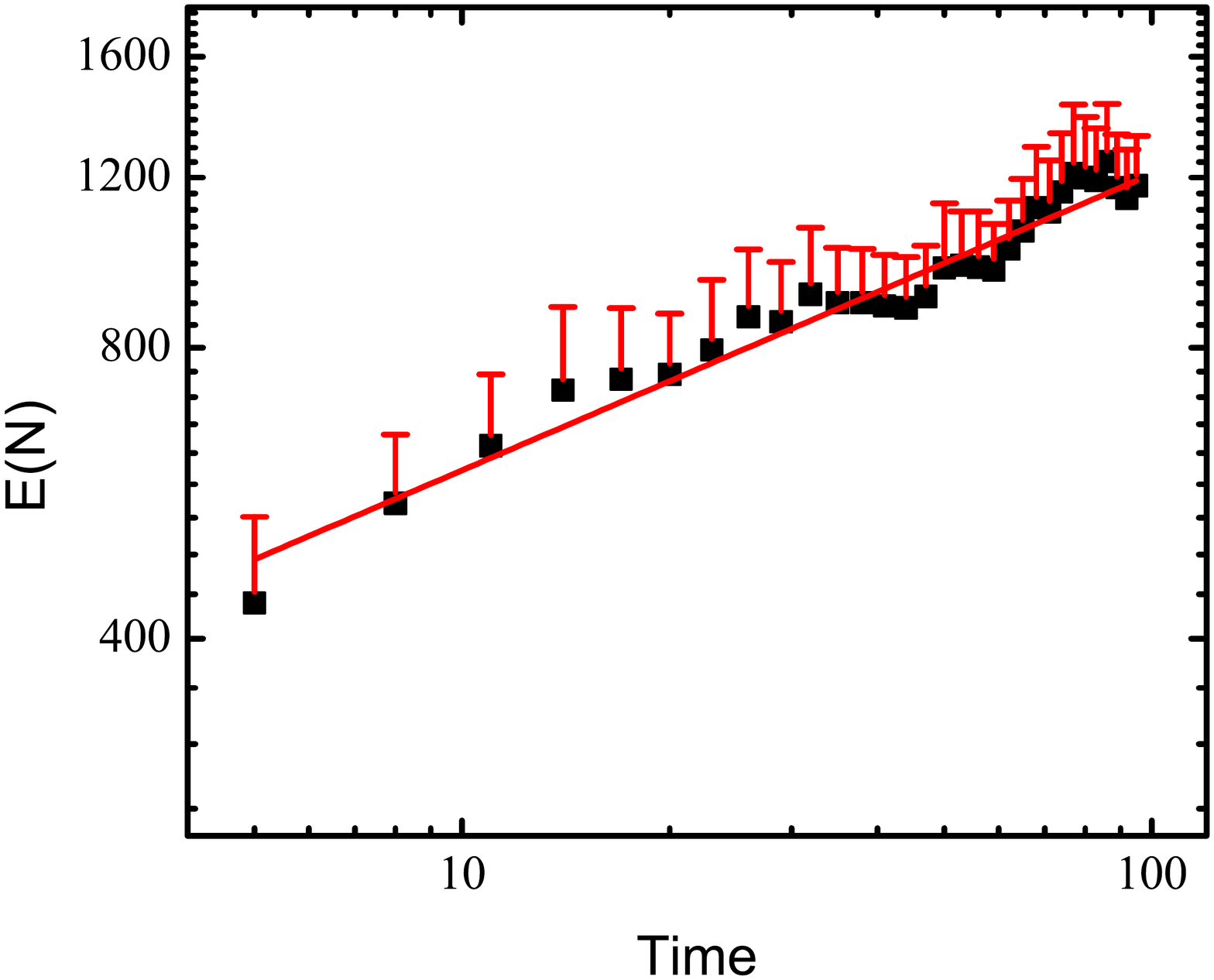}}
\subfigure[Reddit dN-dT] {\includegraphics[width=4.0cm]{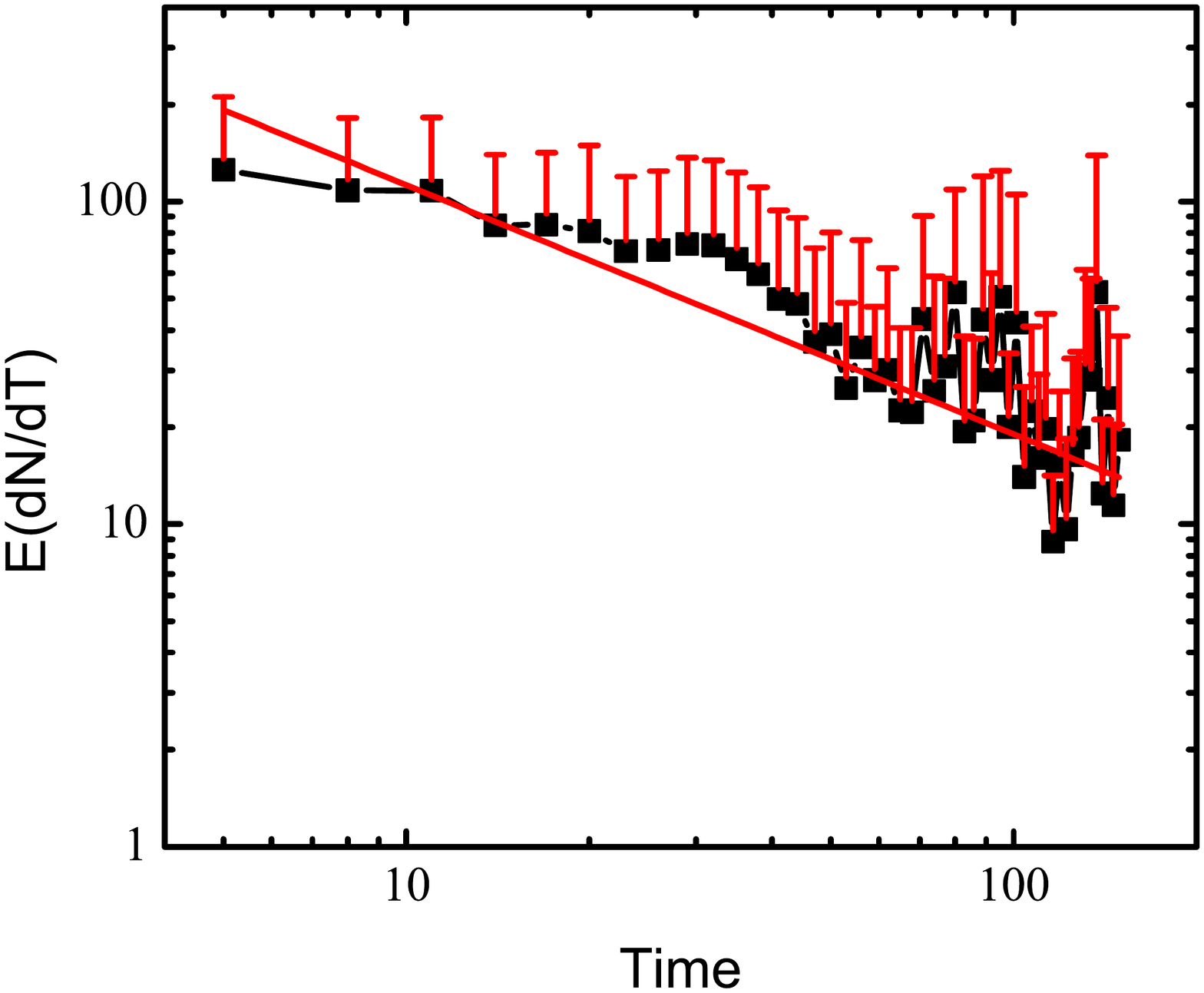}}
\subfigure[Reddit N-dT]{\includegraphics[width=4.0cm]{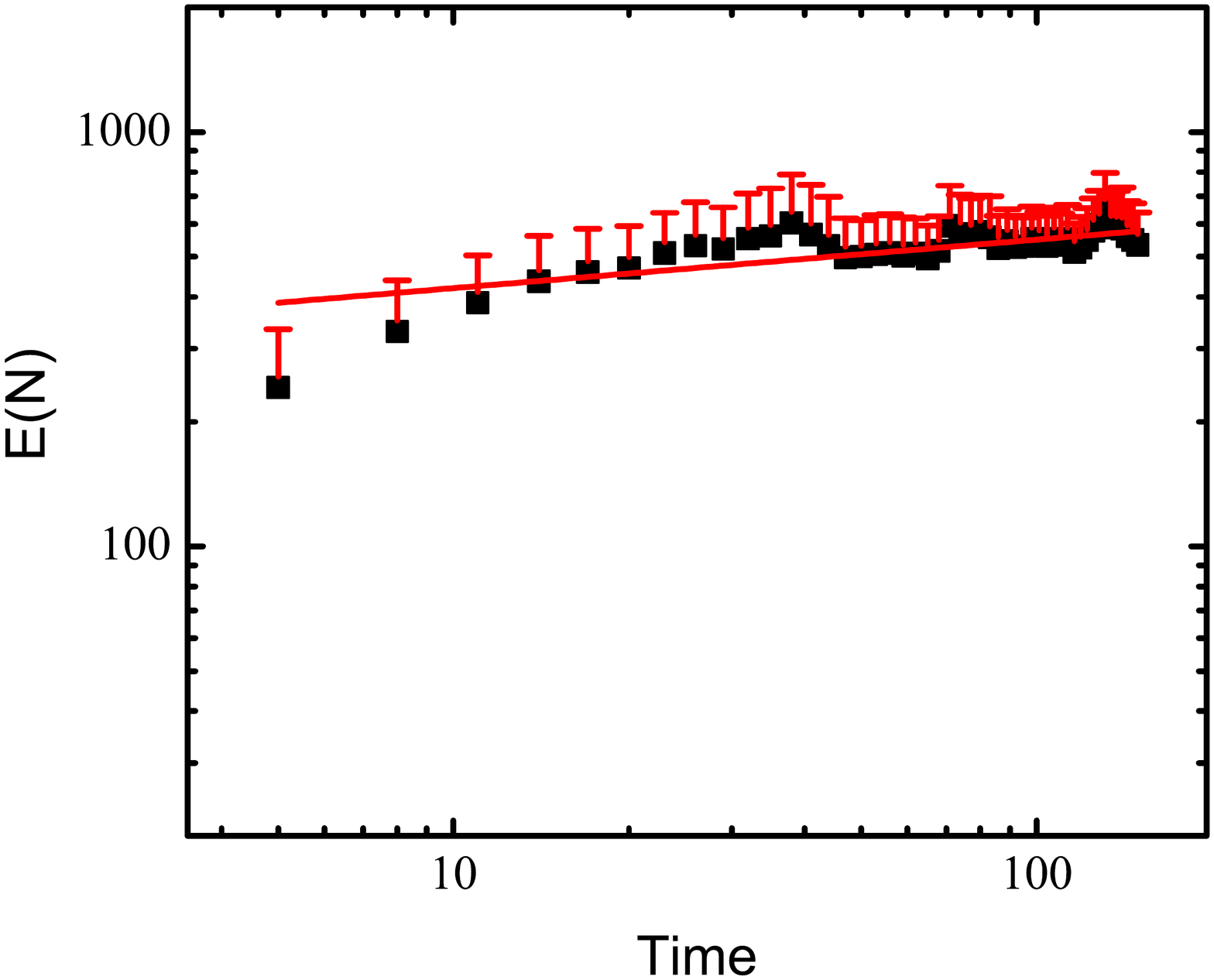}}
 \caption{Comment growth dynamics on Digg in log-log scale. (a) dN/dT as a function of T, the red line in the figure shows the linear fit of data in log-log scale, with a slope of $-0.5483$ and a standard error of $0.3809$. (b) N as a function of T,  the red line in the figure shows the linear fit of data in log-log scale, with a slope of $0.3066$ and a standard error of $0.046$.  Comment growth dynamics on Reddit in log-log scale. (c) dN/dT as a function of T, the red line in the figure shows the linear fit of data in log-log scale, with a slope of $-0.7720$ and a standard error of $0.2454$. (d) N as a function of T,  the red line in the figure shows the linear fit of data in log-log scale, with a slope of $0.1172$ and a standard error of $0.0057$.}
 \label{fig:growth_dynamics}
\end{figure}

Based on results form Section 4, we assume that users check the websites' ``new'' columns regularly to discover topics to read, share and comment on. Before the inflection point, topics can be discovered by these users. In our model, we use $t_0$ to denote the time point when the topic is created. We use $t$ to denote the time passed since the creation of the topic, and $T$ to denote the exposure duration for a topic. We use $N(t)$ to denote the cumulative count of user comments on a topic, or the size of conversation. The waiting times of two consecutive comments from a user follows a upper truncated Pareto distribution. Here, we simplify the problem by assuming that users share the same microscopic behaviors, i.e. the waiting times for different users come from the same distribution. In doing so, we are able to model the process of $M$ users as $M$ independent concurrent counting process, so it is sufficient to consider the case of one individual user.  For that user, the waiting times between two comment is an independent and identical variable. The counting process of that specific user thus forms a renewal process, as depicted in Figure~\ref{fig:renewal_process_demo}. Here, we let $x_i$ denote the inter-arrival time of the $ith$ comment from the user, $Y(t_0)$ denote the time from $t_0$ until the next renewal, $A(t_0)$ denote the time from $t_0$ since the last renewal. If the waiting times of each users' comments follows an independent and identical upper-truncated Pareto distribution, which can be written as
\begin{equation}
f(x) = \frac{c}{{{a^{ - c}} - {b^{ - c}}}}{x^{ - c - 1}}.
\end{equation}
Here, $a$ is the lower bound of the time interval required for a user to post another comment, and $b$ is the upper bound of the Pareto distribution stemming from its finite size effect. The cumulative distribution of the truncated power-law takes form of $F(x) = \frac{{{a^{ - c}} - {x^{ - c}}}}{{{a^{ - c}} - {b^{ - c}}}}$, for $a \le x \le b$. We have
%\begin{equation}
%\overline {F(x)}  = 1 - F(x) = \left\{ {\begin{array}{*{20}{c}}
%{1,\begin{array}{*{20}{c}}
%{}&{}
%\end{array}x < a}\\
%{\frac{{{x^{ - c}} - {b^{ - c}}}}{{{a^{ - c}} - {b^{ - c}}}},a \le x \le b}\\
%{0,\begin{array}{*{20}{c}}
%{}&{}
%\end{array}x \ge b}
%\end{array}} \right.
%\label{eq:cumulative_dist}
%\end{equation} 
For an individual user, the quantity of interest is the probability that a comment happens at $Y(t_0)$ on the specified topic, when $t_0 \to \infty$. To derive this, we begin with the probability of the user commenting on any of the existing topics in the column at time $Y(t_0)$. Note that if the inter-arrival time is independent and identically distributed, $Y(t_0)$ and $A(t_0)$ form an alternating renewal process. From results of the key renewal theorem~\cite{C75}, we have
\begin{equation}
\mathop {\lim }\limits_{t_0 \to \infty } P\{ Y(t_0) \le y\}  = \frac{1}{\mu }\int_0^y {\overline {F(x)} dx} ,\begin{array}{*{10}{c}}
{}&{}
\end{array}(0 \le y \le b).
\label{eq:yt_dist}
\end{equation}
In this equation, $\mu = \frac{c}{{c - 1}}\frac{{{a^{ - c + 1}} - {b^{ - c + 1}}}}{{{a^{ - c}} - {b^{ - c}}}}$ is the expectation of random variable $x$. $A(t_0)$ and $Y(t_0)$ do not have to share the same distribution, due to the impact of the lower bound $a$. By taking Equation~\ref{eq:cumulative_dist} back to Equation~\ref{eq:yt_dist}, we obtain the probability of user comment on any of the items at $Y(t_0)$,
\begin{equation}
P\{ Y(t_0) = y\}  = \frac{{\partial {P(Y(t_0)\le y)}}}{{\partial y}} \sim {y^{ - c}}.
\end{equation}
Based on our assumptions, users can choose to comment on one topic from the column at a time. So at $t_{0}$, the user may choose to comment on any of the existing topics that are still in the column. Neglecting all other factors, if we assign a fixed probability $\alpha$ for the user to choose the specified topic from all topics in the column, the size of conversation scales with $N(t) \sim \alpha t^{-c+1}$. One insight of this equation is that the probability of one more additional comment adding to the topic inversely scales with time. The interestingness measurement $\alpha$ can be assigned different values to different topics. To derive the most common properties of conversation growth dynamics, we fix $\alpha$ over topics in our model. 

Thus far, we have derived the growth of conversation size without considering the resonating nature and the social propagation part of on-line conversations. Now we take these important characteristics into consideration by writing $\alpha$ as $\alpha(N(t))$. The reason that $\alpha$ is a function of $N(t)$ comes from existing works in information cascades and social influence. The more popular a topic is, the more likely that a user comment on it or come back to comment again. Given that $N$ scales with $t$, we assume $\alpha = \gamma t^{c_0}$, $\gamma$ is a constant factor. $c_0$ is a positive exponent measuring the combined impacts of factors such as resonance and social influence. In the extreme of $c_0 = 0$, $\alpha$ would be a constant, when there is no other impacts such as social influence. Now we combine existing two parts together to derive the dynamics of conversation size growth. Noting that the expected number of increment of comments at time point $t$ would be proportional to $\gamma t^{c_0}Mt^{-c}$. The total number of comments for a given topic, $N$, grows like
\begin{equation}
\frac{{dN(t)}}{{dt}} = \gamma Mt{^{ - c+c_{0}}}.
\end{equation}
Thus, $N(t) \sim t{^{ - c + c_{0} + 1}}$, i.e. $ln(N(t))$ scales linearly with $ln(t)$. To confirm this derivation, we compare this result with the empirically measured growth of conversation size. As shown in Figure~\ref{fig:growth_dynamics}, the plot in log-log scale shows the expected scaling relationship between time and number of increments. In the figures, the red line in the figure shows the linear fit of data in log-log scale. For Digg, linear fitting gives a slope of $-0.5483$ and a standard error of $0.3809$ as shown in Figure~\ref{fig:growth_dynamics} (a) and a slope of $0.3066$ with a standard error of $0.046$ for Figure~\ref{fig:growth_dynamics} (b). For Reddit, linear fitting yields a slope of $-0.7720$ and a standard error of $0.2454$ for Figure~\ref{fig:growth_dynamics} (c) and a slope of $0.1172$ with a standard error of $0.0057$ for Figure~\ref{fig:growth_dynamics} (d). The two estimated exponent values obtained from two fittings have a difference around one for both Digg and Reddit, which result agrees well with the relationship of $N(t) \sim t{^{ - c + c_{0} + 1}}$ and $\frac{dN(t)}{dt} \sim t{^{ - c + c_{0}}}$ from our model. One point worth mentioning here is that, in some of existing works about attention dynamics, the above derived relationship is used as an assumption upon which the model is built~\cite{F07,L10}. 

Now, we turn our attention to the distribution of conversation sizes when  topics reaches their inflection points, i.e. $N(T)$, based on the observed distribution of $T$. For simplificity, we use $c' = -c + c_{0}+1$, and $\gamma' = \gamma M$ in our derivations,  so that $N(T) = \gamma' T^{c'}$. We then have
\begin{equation}
P(N(T) \le n) = P(\gamma' {T^{c'}} \le n) = {\mathop{\rm P}\nolimits} (T \le {(\frac{n}{\gamma' })^{\frac{1}{{c'}}}}).
\label{eq:beforetime}
\end{equation}
The actual form of the cumulative distribution depends on the distribution of exposure durations, which is website specific as discussed earlier. We now discussed in more detail the impact that different exposure duration distributions have on the empirically measured conversation size distributions. We look at two general cases of exposure duration: (i) exponential distribution and (ii) Pareto distribution.

\subsection{Exponential Exposure Duration}
\begin{figure}[htl]
 \centering
\hspace{-10pt}
 \subfigure[Density Plot]{\includegraphics[width=4.4cm]{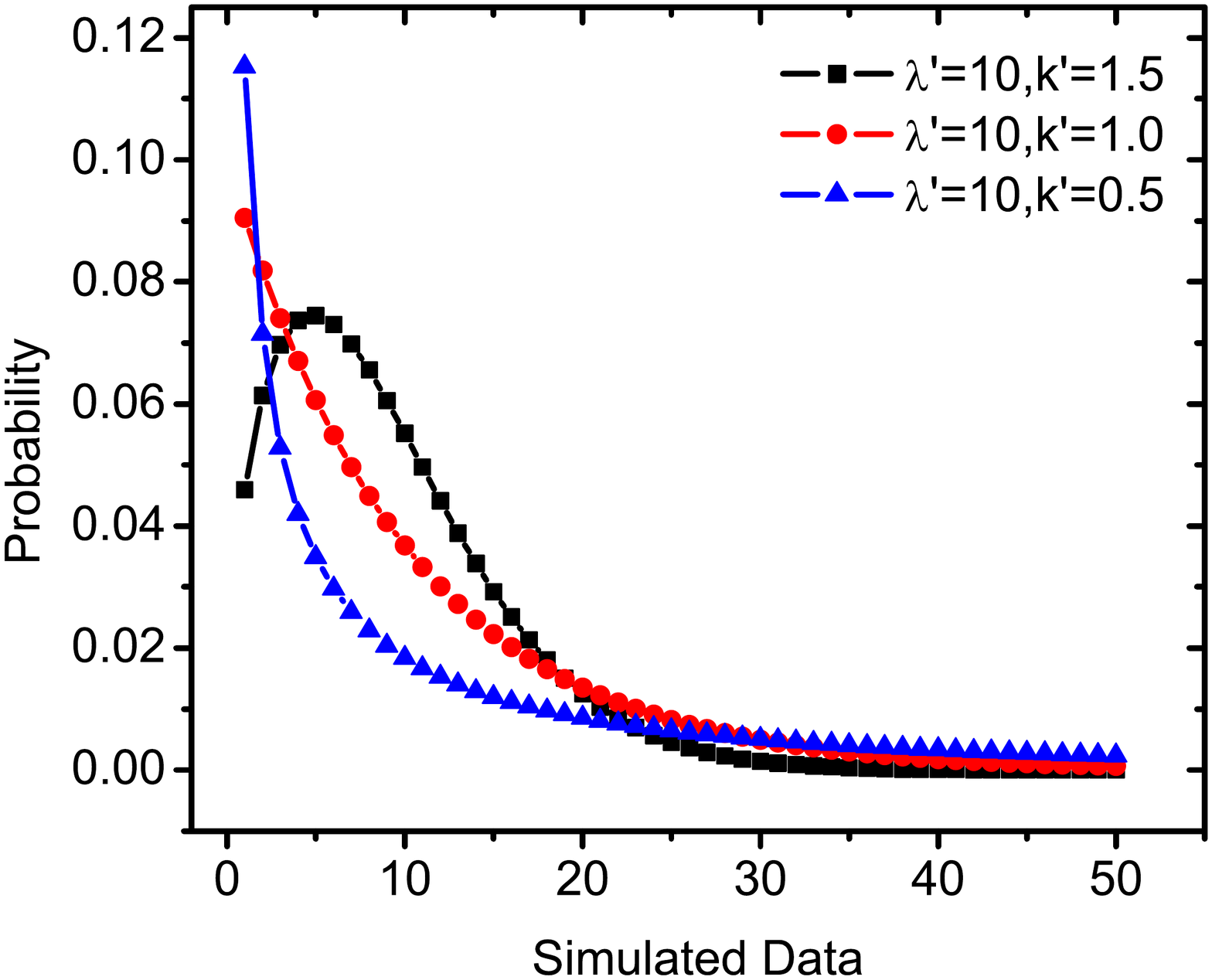}}
\hspace{-10pt}
\subfigure[CCDF Plot]{\includegraphics[width=4.4cm]{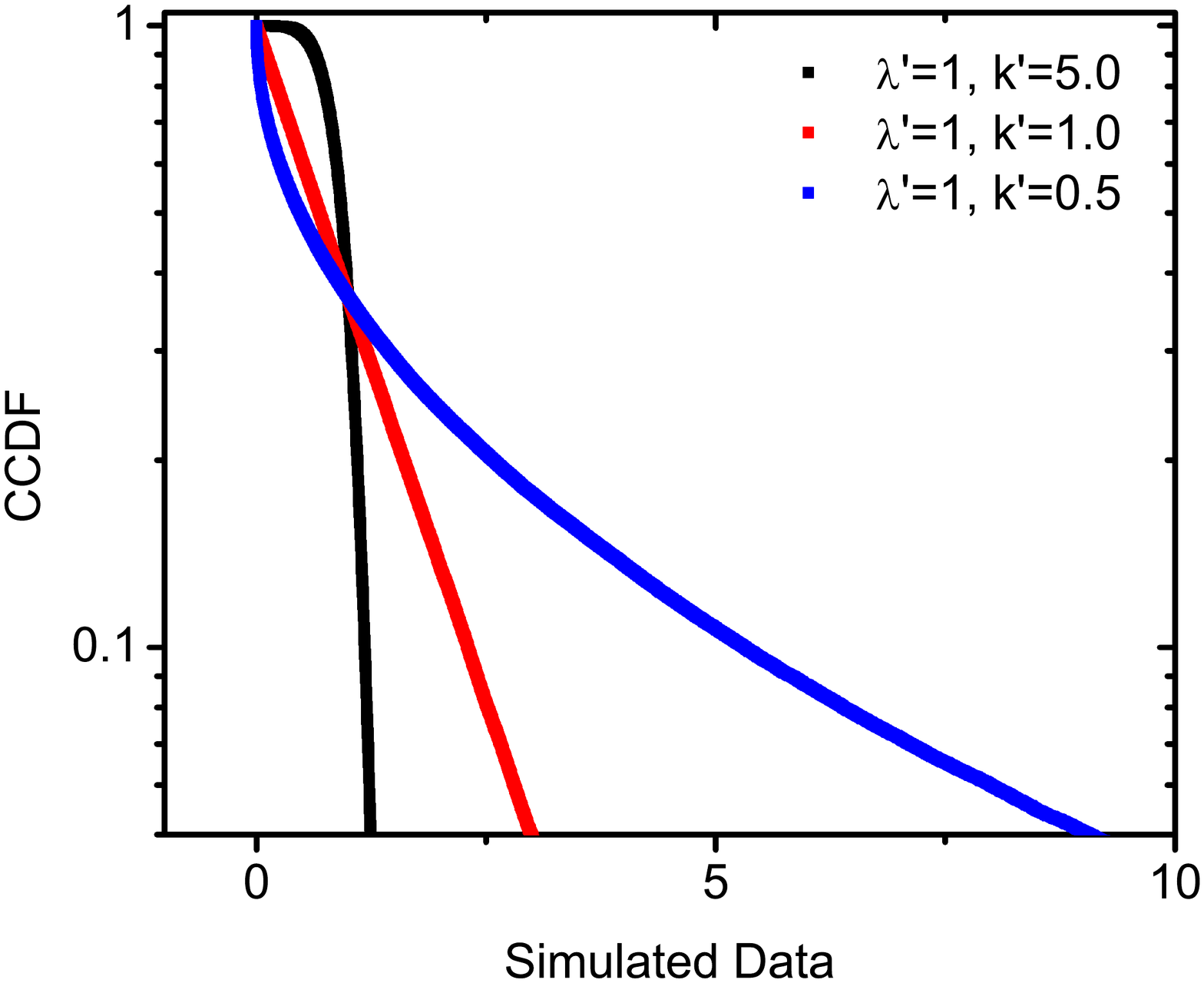}}
\hspace{-10pt}
 \caption{ (a) Density plot of simulated size for the situation of exponentially distributed exposure duration, when $\lambda' = 10$ and (i)$k' = 1.5$, (ii)$k' = 1.0$, (iii) $k' = 0.5$. (b) CCDF plot of simulated size, when $\lambda' = 1$ and (i)$k' = 5.0$, (ii)$k' = 1.0$, (iii) $k' = 0.5$.}
 \label{fig:simulated_weibull_size}
\end{figure}

For an exponentially distributed $T$ with rate parameter $\lambda$ as measured in Digg and Reddit, its cumulative distribution has a form of $P(T \le x ) = 1 - e^{-\lambda x}$, So take this back to Equation~\ref{eq:beforetime}, we have
\begin{equation}
{\mathop{\rm P}\nolimits} (N(T) \le n) = 1 - {e^{ - \lambda {{(\frac{n}{\gamma'})}^{\frac{1}{{c'}}}}}}. 
\end{equation}
By taking the derivative of this equation, we arrive at the distribution of $N(T)$ taking the form of:
\begin{equation}
{\mathop{\rm P}\nolimits} (N(T) = n) = \frac{\lambda }{{c'\gamma' }}{(\frac{n}{\gamma' })^{\frac{1}{{c'}} - 1}}{e^{ - \lambda {{(\frac{n}{\gamma'})}^{\frac{1}{{c'}}}}}}.
\end{equation}
This is actually a Weibull distribution with its shape parameter $k'$ equals $\frac{1}{c'}$ and scale parameter $\lambda'$ equals $\frac{\gamma'}{\lambda^{c'}}$. Interestingly, the tail of the distribution scales as $e^{ - \lambda {{(\frac{n}{\gamma' })}^{\frac{1}{c'}}}}$. So the distribution has following properties:
\begin{itemize}
\item{{\bf Case 1: ($k' = \frac{1}{c'} < 1$)} In this case, when the social influence factor has a stronger impact than the decay factor, $c' > 1$, the shape factor is smaller than one. So
\begin{equation}
\mathop {\lim }\limits_{n \to \infty } {e^n}P(N > n) = \infty, 
\end{equation} 
which results in a heavy tailed distribution of conversation size.  } 

\item{{\bf Case 2: ($k' = \frac{1}{c'} > 1$)} In this case, the tail decays faster than an exponential distribution. The distribution would appear to be light-tailed.}

\item{{\bf Case 3: ($k' = \frac{1}{c'} = 1$)} This is the case when the size distribution has an exponential distribution, which is corresponding to the red line in  Figure~\ref{fig:simulated_weibull_size} (a) and (b) .}
\end{itemize}

Thus for the case of exponentially distributed exposure duration, both heavy tailed and non-heavy tailed distributions can appear. The actual form of the distributions is determined by the factor $c_0$. If the social propagation dominates, there is a good chance that one would observe extremely large comment threads. Figure~\ref{fig:simulated_weibull_size} (a) demonstrates the simulated density plot under the three cases of heavy-tailed, exponential and light-tailed size distribution. And Figure~\ref{fig:simulated_weibull_size} (b) shows the complementary cumulative distribution function(CCDF) Plot in a semi-log scale of above three cases. If the tail is not exponentially bounded, the CCDF curve will lie above the straight lin,e as seen in the blue one in Figure~\ref{fig:simulated_weibull_size} (b).

\subsection{Pareto Exposure Duration}
Next, we investigate the situation when the exposure duration follows a Pareto distribution, as measured in Epinions. We have 
\begin{equation}
P(T < x) = 1 - {(\frac{x}{{{T_{\min }}}})^{ - \alpha }}, x > T_{\min}.
\end{equation}
By taking this back to Equation~\ref{eq:beforetime}, we have
\begin{equation}
P(N(T) \le n) = P(\gamma'{T^{c'}} \le n) = 1 - T_{\min}^{-\alpha}{(\frac{n}{{\gamma'}})^{ - \frac{\alpha }{{c'}}}}.
\end{equation}
Taking the derivative on $n$, the size distribution has the form of:
\begin{equation}
P(N(T) = n) \sim \frac{\alpha }{{c'}}{(\frac{n}{{\gamma'}})^{ - \frac{\alpha }{{c'}} - 1}},
\end{equation}
which is a Pareto distribution. So the conversation size of topics with a Pareto exposure duration has a heavy-tail.

\hspace{10pt}

From the above analysis of conversation size distribution under different exposure durations, we can see that the discrepancies in the reported size distributions stem from the hidden algorithms that websites employ for deciding which new topics to display on their websites. By separating these artificial factors from user behavior, we show that there are actually underlying mechanisms in common for different social media websites. This explains why different categories of distributions (heavy tailed and non-heavy tailed) are observed in existing studies~\cite{M06, K10, K11,G08,O08,T09}. The model can be adapted to other empirically measured exposure durations. For instance, for a log-normal distribution of exposure duration, a log-normal size distribution is expected from our model. Due to the space limitations, we omit the derivations here. We compare the predictions of this model with empirical measurements in Section 6.

\subsection{Structure of Conversation}
\begin{figure}[htl]
 \centering
 \includegraphics[width=4.8cm,height=3.3cm]{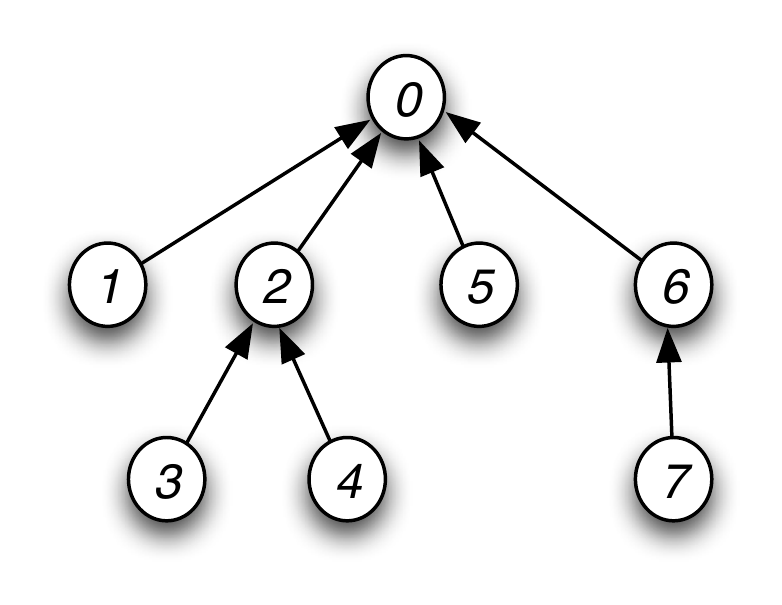}
\vspace{-5pt}
 \caption{Structure of a sample conversation thread from Digg. Node $0$ is the initial post of the topic.}
 \label{fig:comments_topology}
\end{figure}

Another interesting characteristics of on-line conversations is the interactive nature of comments. For example, when a new comment is added to the thread, it is following either the original post or one of the existing comments, so that the comments form a directed graph with each comment as a node. Figure~\ref{fig:comments_topology} shows one  such example based on a sample thread of comments from Digg. The node $0$ is the original post of topic, and the others nodes represent the following comments. Node $2$ has a in-degree of $2$ in this graph.  Within such an information network formed by user comments, one of the most important properties is the in-degree distributionof the associated directed graph. We now show that by adding to our dynamic model for conversations a Yule process~\cite{Y24,J09}, one can derive the in-degree distribution of the network. The Yule process assumes that each new comment added into the thread follows one simple rule: it is added to one of the existing comments with a probability proportional to their existing in-degree. Notice that there is also a probability to comment on one with zero in-degree, denoted by $\delta_0$. We use $k_j(t)$ to denote the in-degree at time $t$ for comments that are created at time point $j$. Based on our model, the number of new comments added into the thread at time $t$ is proportional to $t^{c'-1}$.  The cumulative count of comments up to time $t$ is proportional to ${t^{c'}}$. The sum of in-degree would thus be equal to $(1 + {\delta _0})\gamma' {t^{c'}}$. Based on our assumptions on Yule process, the probability of a new comment attaching to an existing comment with $k_i$ in-degree would be $\frac{{({k_i(t)} + {\delta _0})}}{{(1 + {\delta _0})\gamma' {t^{c'}}}}$. Since there are ${c'\gamma' {t^{c' - 1}}}$ new comments added at $t$, the growth dynamics of $k_j(t)$ can be written as
\begin{equation}
\frac{{\partial {k_i(t)}}}{{\partial t}} = \frac{{({k_i(t)} + {\delta _0})c'\gamma' {t^{c' - 1}}}}{{(1 + {\delta _0})\gamma' {t^{c'}}}}.
\end{equation}
After integrating this equation and taking into account the initial condition $k_i(i)=0$, we have
\begin{equation}
\ln (\frac{{{k_i(t)} + {\delta _0}}}{{{\delta _0}}}) = \frac{{c'}}{{1 + {\delta _0}}}\ln (\frac{t}{{{t_i}}}).
\end{equation}
The in-degree of node $i$ at the inflection point equals to
\begin{equation}
{k_i}(T) = {\delta _0}[{(\frac{T}{{{t_i}}})^{\frac{{c'}}{{1 + {\delta _0}}}}} - 1].
\end{equation}
So now we have
\begin{equation}
P({k_i(T)} < l) = P({\delta _0}[{(\frac{T}{{{t_i}}})^{\frac{{c'}}{{1 + {\delta _0}}}}} - 1] < l) = P({t_i} >T {[1 + \frac{l}{{{\delta _0}}}]^{ - \frac{{1 + {\delta _0}}}{{c'}}}})
\label{eq:indegree_cumulative}
\end{equation}
For a randomly chosen node, $P(t_i>t)=1-\frac{t^{c'}}{T^{c'}}$, so  Equation~\ref{eq:indegree_cumulative} becomes
\begin{equation}
P({k_i}(T) < l) = 1 - {[\frac{l}{{{\delta _0}}} + 1]^{ - 1 - {\delta _0}}}.
\end{equation}
By taking the derivative of this equatio we see that the distribution of  in-degrees scales with $[\frac{l}{{{\delta _0}}} + 1]^{ -2 - {\delta _0}}$, which amounts to a Pareto scaling. An interesting consequence from the above derivation is that the in-degree distribution within a conversation network is independent of the distribution of exposure durations. That is to say, despite of the different hidden algorithms used by websites, the structure within a conversation is determined by user behavior and  universal. This result explains why existing studies observe the same Pareto scaling of in-degree distribution within conversation threads in different social media sites~\cite{M06, K10, K11,G08,O08,T09}.

\section{Empirical Observations}

\begin{figure*}[thl]
 \centering
  \subfigure[Digg]{\includegraphics[width=5.5cm]{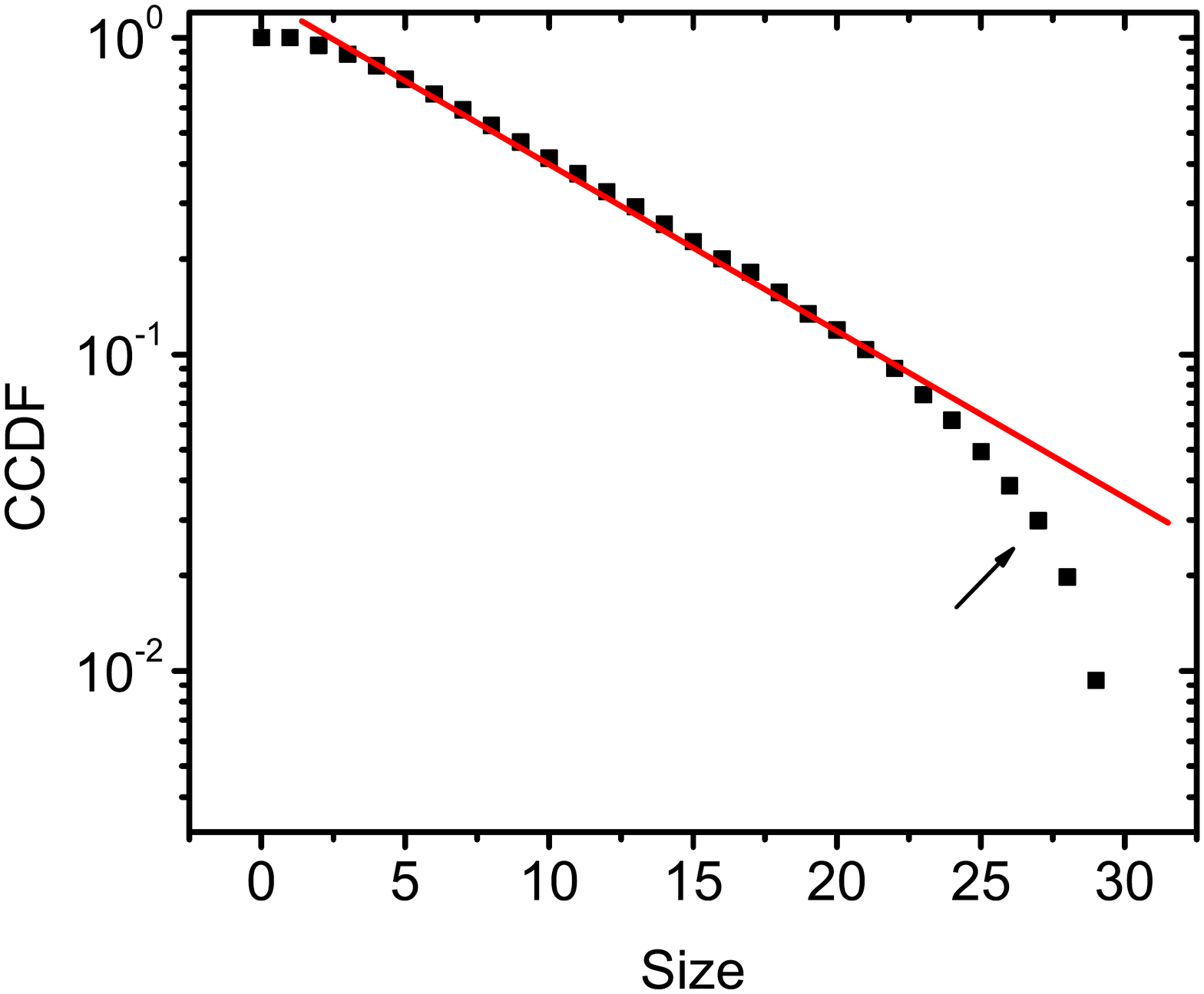}}
\hspace{-10pt}
  \subfigure[Reddit] {\includegraphics[width=5.5cm]{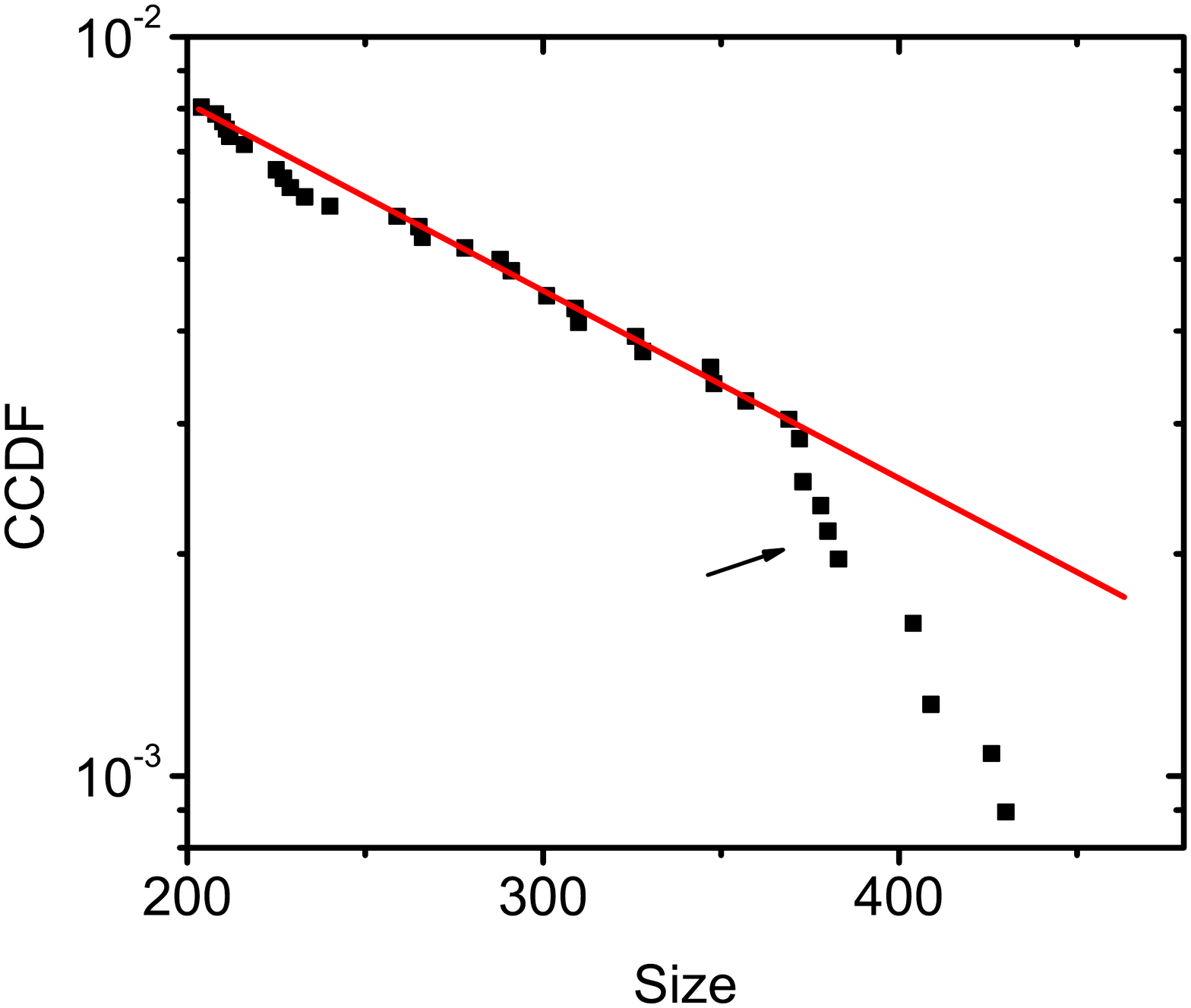}}
\hspace{-10pt}
  \subfigure[Epinions]{\includegraphics[width=5.5cm]{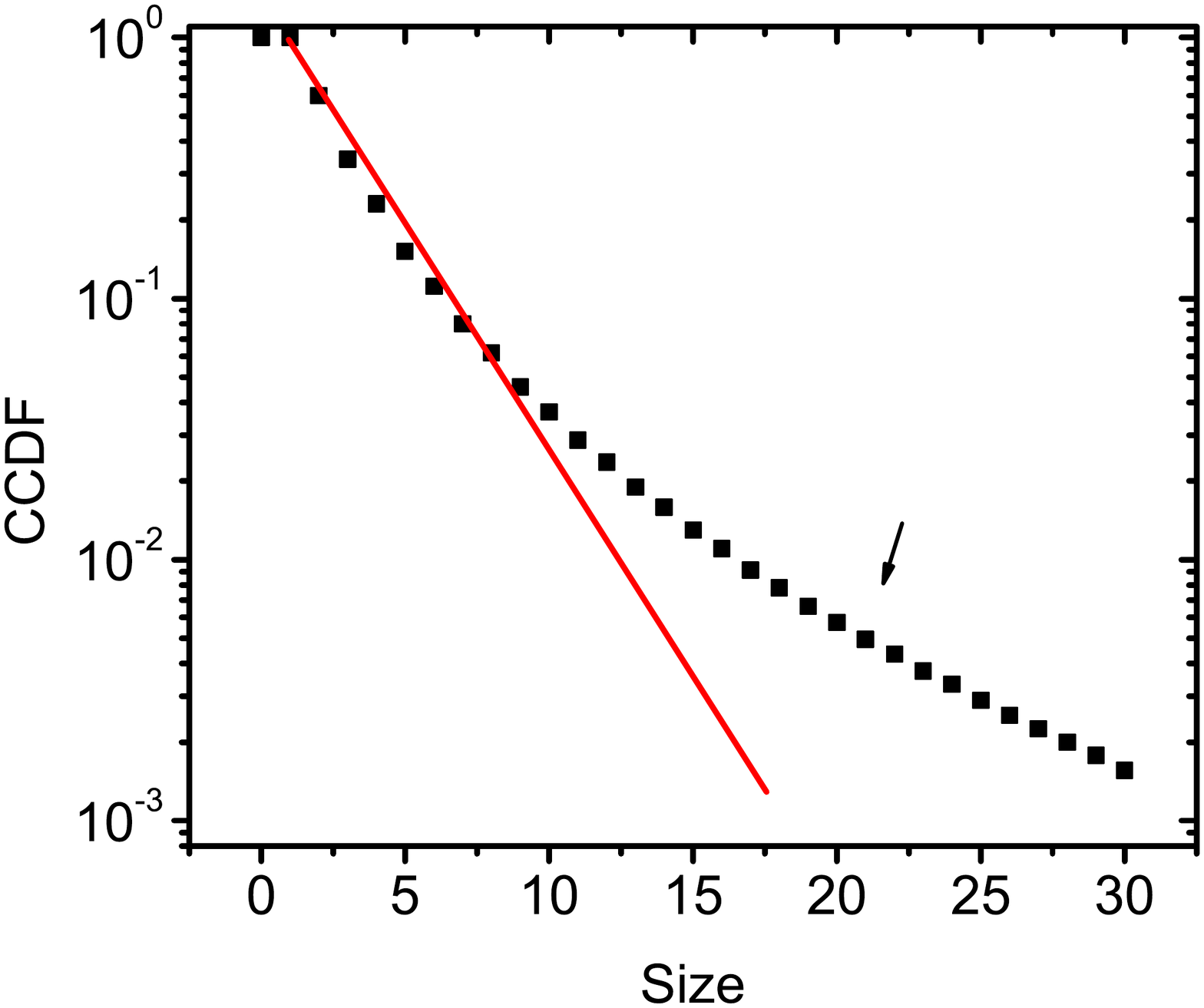}} \\
\vspace{5pt}
\begin{tabular}{|l|c|c|c|}
\hline & Digg & Reddit & Epinions\\
\hline Topic Exposure Duration Distribution & Exponential & Exponential & Pareto \\
\hline Predicted Size Distribution & Weibull (Light Tailed) &  Weibull (Light Tailed) & Pareto \\
\hline Observed Size Distribution  & Weibull (Light Tailed) & Light Tailed & Pareto \\
\hline
\end{tabular}\\\vspace{10pt}
{(d) Summarization of the size distribution in different websites.}
 \caption{ (a) Digg, (b) Reddit and (c) Epinions, the CCDF Plot in semi-log scale for three social media websites. If the density tail(black squares in the plot) is above the red exponential line, then it is a heavy-tailed distribution, otherwise not. (d) Summarization of predicted and measured tail properties in different websites.}
 \label{fig:size_ccdf}
\end{figure*}

In the last section we modeled the process of conversation growth and predicted that the distribution of conversation sizes is determined by several factors including the exposure duration, the users' commenting behavior, the social propagation and resonating factors. We also demonstrated that a universal Pareto in-degree distribution is expected for each comment. In this section, we compare these results with empirical observations. 

 \begin{figure}[htl]
 \centering
\hspace{-10pt}
 \subfigure[Digg]{\includegraphics[width=4.4cm]{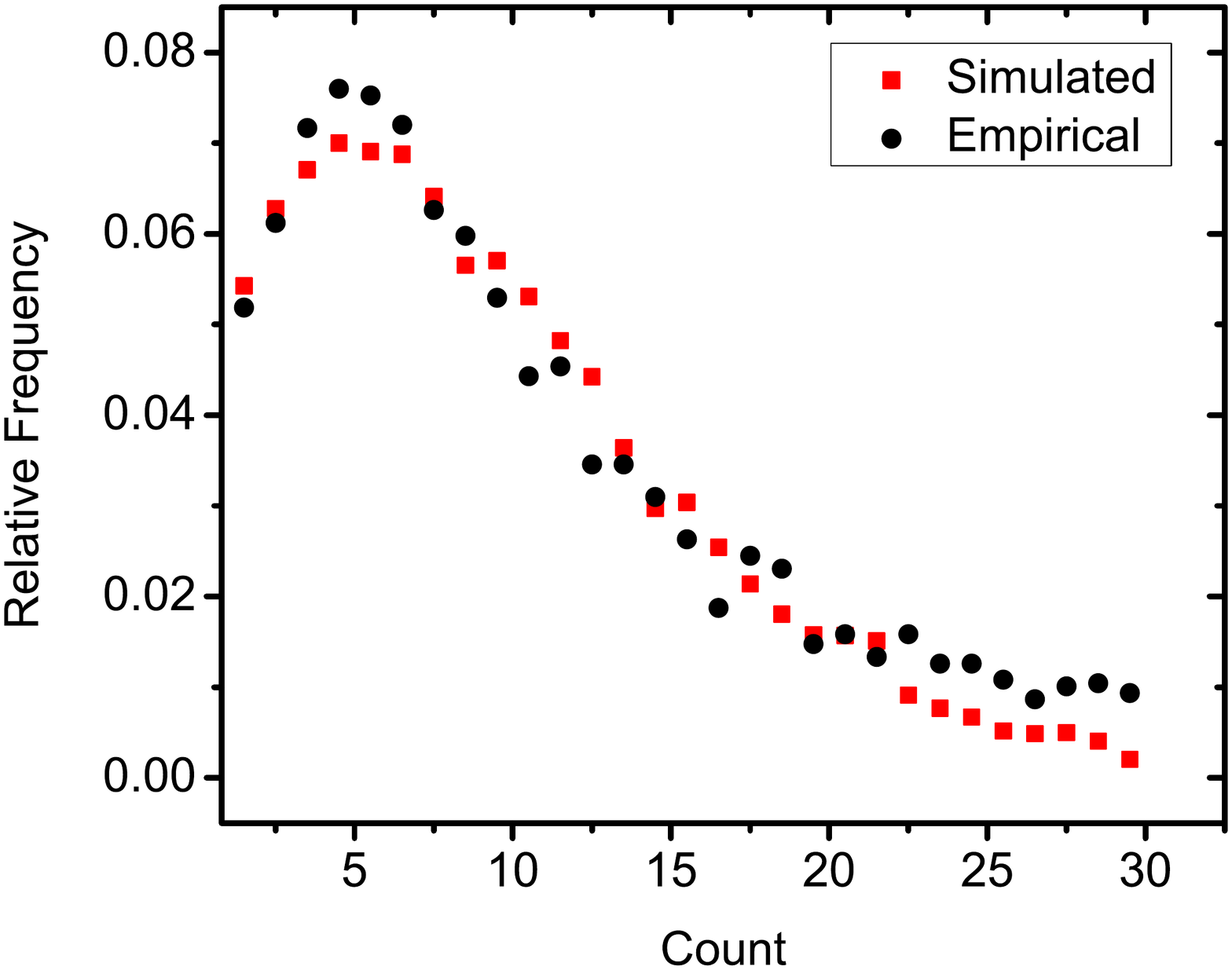}}
\hspace{-10pt}
 \subfigure[Epinions]{\includegraphics[width=4.4cm]{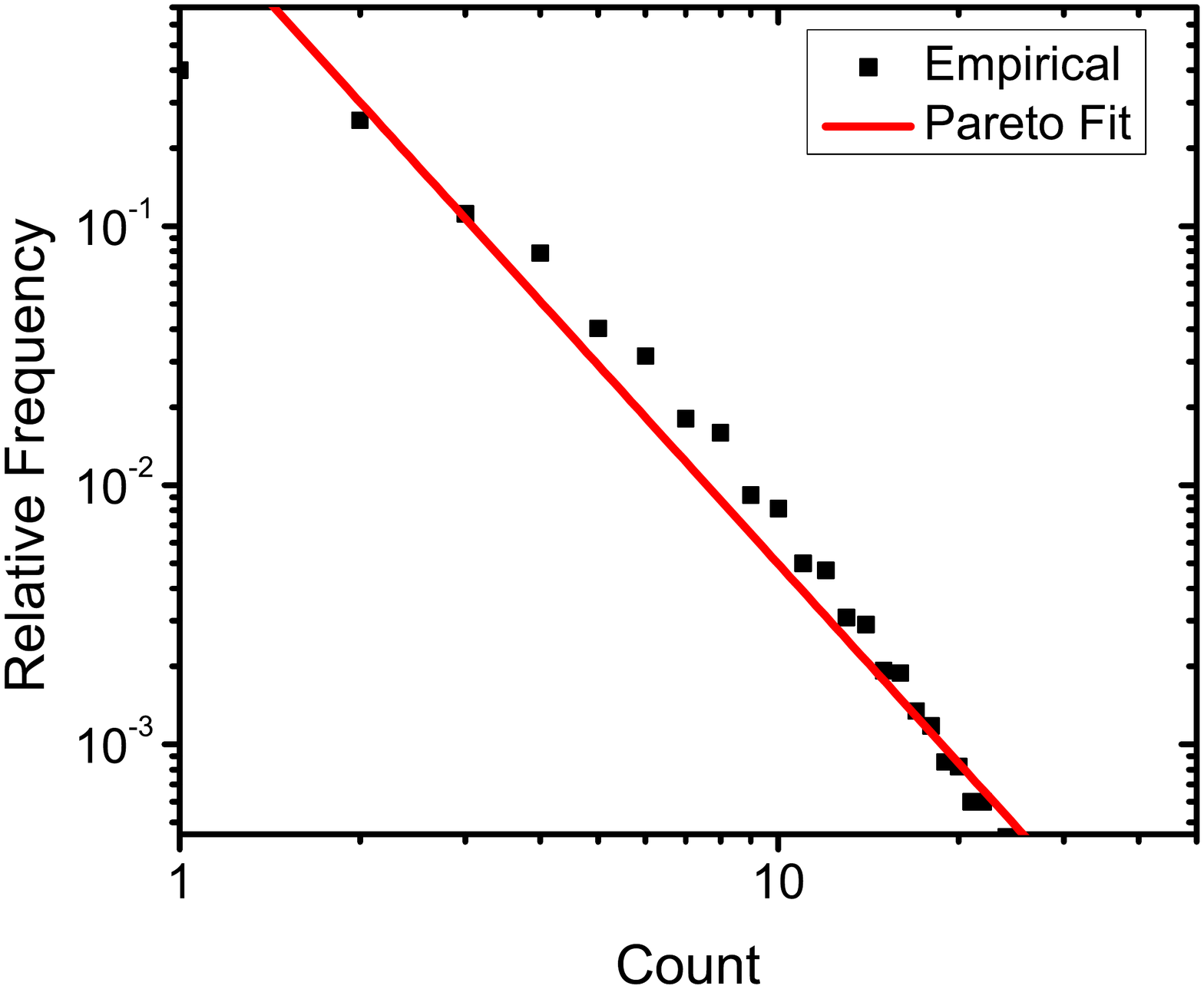}}
\hspace{-10pt}
 \caption{The density plot of conversation size on (a) Digg and (b) Epinions. }
 \label{fig:simulated_comment_size}
\end{figure}

\begin{figure}[htl]
 \centering
  \subfigure[Weibull Plot] {\includegraphics[width=4cm]{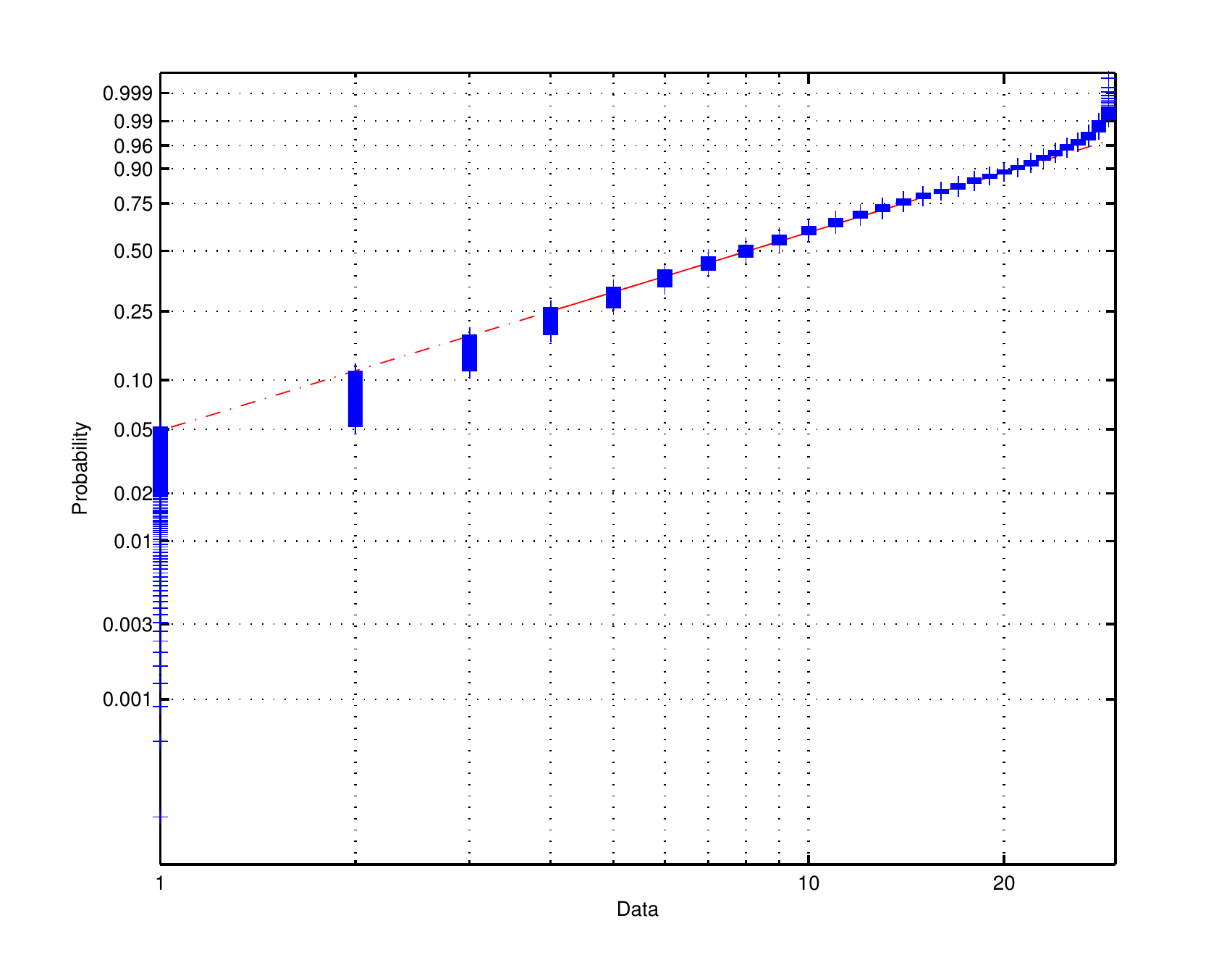}}
  \subfigure[QQ-Plot]{\includegraphics[width=4cm]{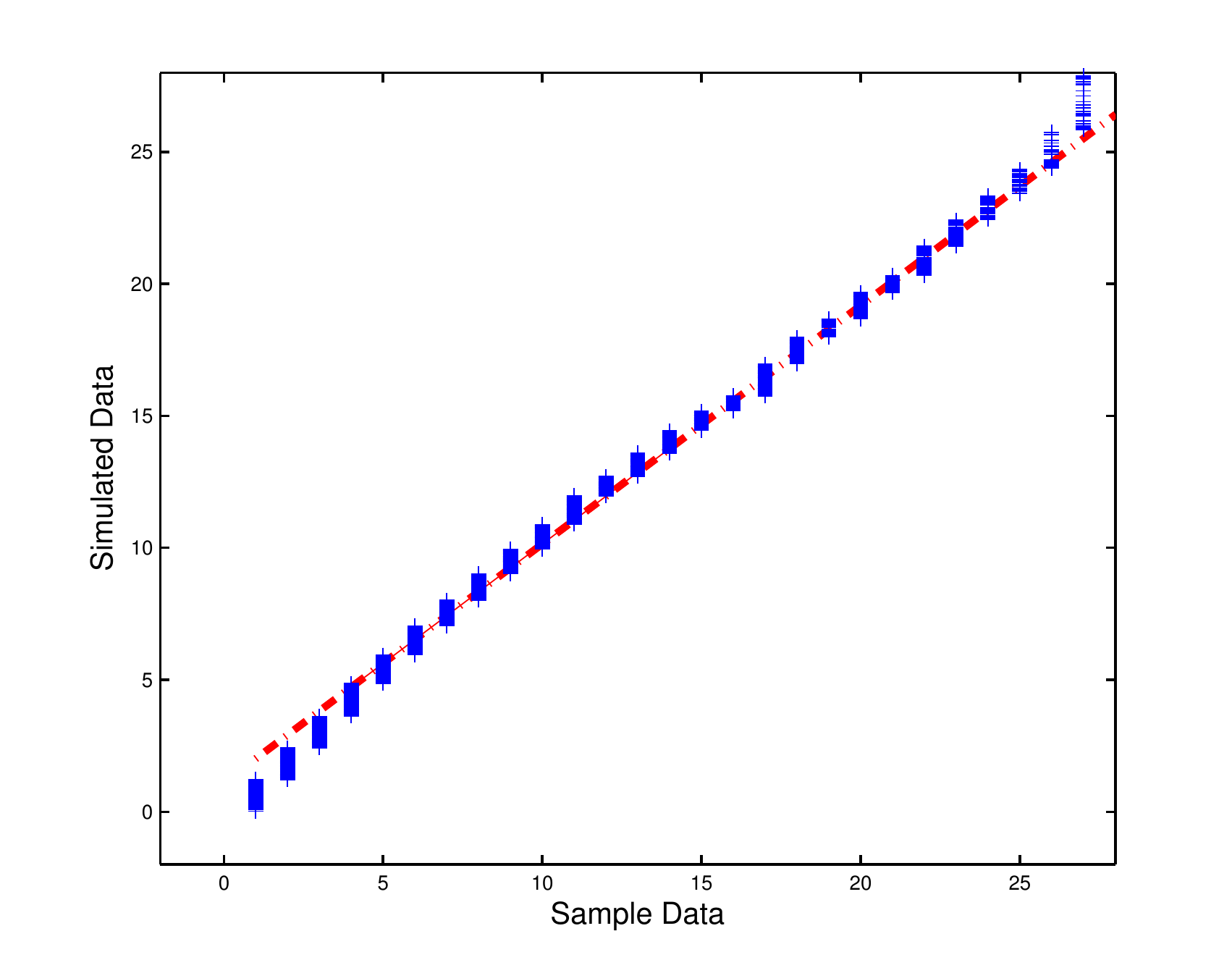}}
 \caption{(a) Weibull Plot of empirically observed conversation size. (b) QQ-Plot of empirically measured and simulated conversation size.}
 \label{fig:weibull_plot}
\end{figure}

First, we compare the size distribution of Digg conversation size with our model. Since  exposure duration within Digg is observed to follow an exponential duration (Figure~\ref{fig:Topic_duration}(a)), and $c'$ is less than one (Figure~\ref{fig:growth_dynamics}), the size distribution of conversation is expected to be described by the second case in Section 5.1, which is a light tailed Weibull distribution. To exclude the impact of stories promoted to front page in the popular column, we filtered out  topics with extreme large comment size. We fit the empirically measured conversation size with Weibull distribution using MLE and estimate that the scale factor equals $10.834$ and the shape factor equals $1.439$. We use the estimated parameters to simulate the conversation size distribution. The density plot of empirical observation and simulated data is as shown in Figure~\ref{fig:simulated_comment_size} (a). In Figure~\ref{fig:weibull_plot} (a), we plot the empirically measured size in a Weibull Plot. And in  Figure~\ref{fig:weibull_plot} (b), we plot the simulated and empirically measured data in a QQ Plot. The straight line in both plots demonstrate that the empirically observed conversation size fits well with the predicted size distribution. From the CCDF Plot in semi-log scale as in Figure~\ref{fig:size_ccdf} (a), we can see that the distribution has a light-tail. Similarly on Reddit, the size distribution is expected to follow a lighted-tailed Weibull distribution, as shown in Figure~\ref{fig:size_ccdf} (b). We also measure the size distribution using dataset from Epinions. Based on Figure~\ref{fig:Topic_duration} (b), the size distribution of Epinions is expected to follow a Pareto distribution from results in Section 5.2. The density plot of size in log-log scale is as shown in Figure~\ref{fig:simulated_comment_size} (b). The Pareto scaling agrees with our model. The distribution is not exponentially bounded as shown in Figure~\ref{fig:size_ccdf} (c). A summary of the tail properties in different social media is shown in Figure~\ref{fig:size_ccdf} (d). The size distributions from Digg and Reddit have a light-tail and that from Epinions a heavy-tail. The derived size distribution from model agrees well with empirical measurements.

\hspace{10pt}

\begin{figure}[htl]
 \centering
  \subfigure[Digg] {\includegraphics[width=4cm]{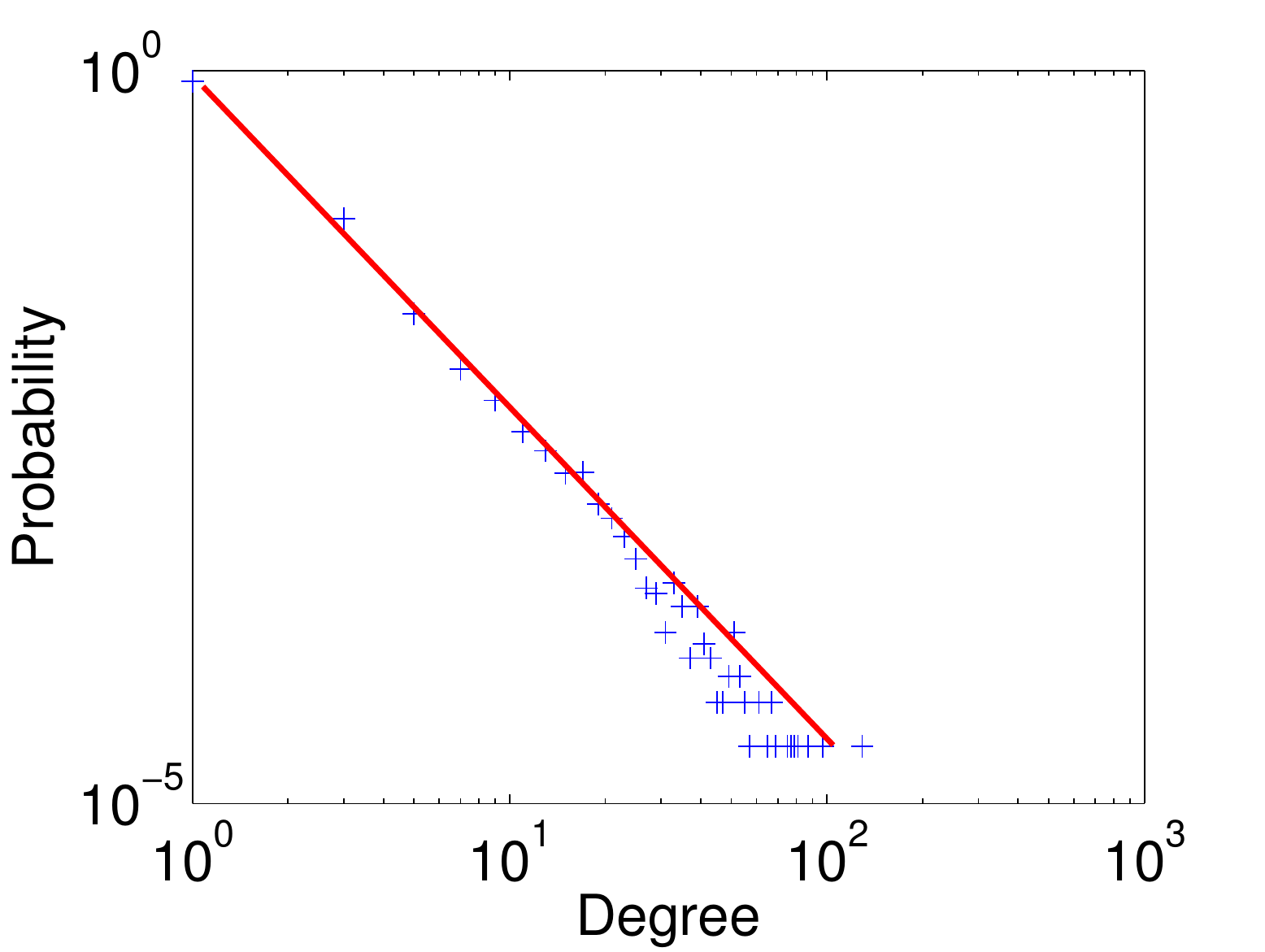}}
  \subfigure[Reddit]{\includegraphics[width=4cm]{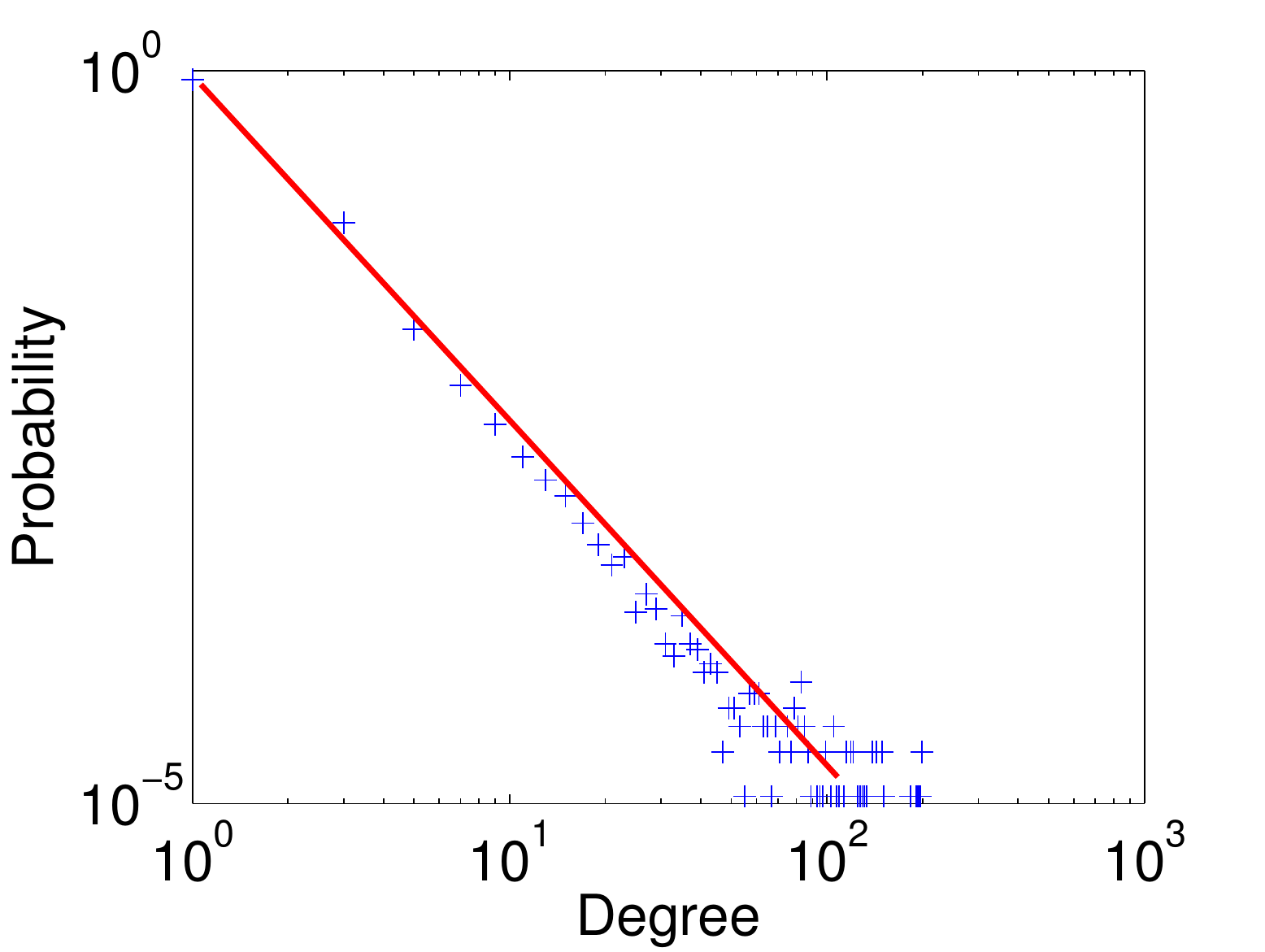}}
 \caption{In-degree size distribution of comments in log-log scale (a) Digg and (b) Reddit. The red straight line in the plots suggests a Pareto distribution.}
 \label{fig:indegree_size_distribution}
\end{figure}

In addition, we compare the derived Pareto scaling of in-degree size distribution for different datasets. Figure~\ref{fig:indegree_size_distribution} shows the density plot of in-degree in a log-log scale of Digg and Reddit. We observed the same scaling in Epinions dataset. The straight line in the figure confirms that the in-degree size follows a Pareto distribution. From the above comparisons, our model quantitatively explains the observed discrepancies of size distribution in different social media, as well as the in-degree size distribution in the information network formed by user comments.  

\section{Conclusion and Future Work}
In this paper, we investigated properties of user conversations in on-line social media. We started from the commenting behavior of individual users and the distribution of exposure duration during which the new topics are displayed to users. Based on these observations, we proposed a general dynamic model for conversation growth. The model successfully explains the reported difference in existing studies from different social media websites. Ee further extended our model with a Yule process to derive the structure of conversations. The results of our model were compared with various empirical measurements, such as the scaling relationship between time and size, the tail properties of size distribution and also the in-degree distribution from different social media sites. Our model provides a powerful framework which can be easily modified and applied to various specific scenarios for studying on-line conversations. Possible refinements of the model may take into considerations of the differences between users, the interestingness of topics, and also the impacts of other featuring mechanisms used by the website. In closing, we note that although the focus in this paper has been on user comments and on-line conversations, the framework of our growth model may be suitable to a wide category of attention dynamics related studies. The wide applicability and the relatively simple assumptions make our model an extremely general one and therefore should provide ample opportunities for future work. 

\section{Acknowledgments}
C. Wang would like to thank HP Labs for financial support.

\bibliographystyle{abbrv}

\end{document}